\documentclass[english,notitlepage,twocolumn,showpacs,pra]{revtex4-1}
\usepackage[T1]{fontenc}
\usepackage[latin9]{inputenc}
\usepackage{amsmath}
\usepackage{amssymb}
\usepackage{esint}
\usepackage{babel}
\begin{document}

\title{Universal relations for a spin-polarized Fermi gas in two dimensions}

\author{Shi-Guo Peng}
\email{pengshiguo@gmail.com}

\affiliation{State Key Laboratory of Magnetic Resonance and Atomic and Molecular
Physics, Wuhan Institute of Physics and Mathematics, Chinese Academy
of Sciences, Wuhan 430071, China}

\date{\today}
\begin{abstract}
We derive the full set of universal relations for spin-polarized Fermi
gases with $p$-wave interaction in two dimensions, simply using the
short-range asymptotic behavior of fermion-pair wave functions. For
$p$-wave interactions, an additional contact related to the effective
range needs to be introduced, besides the one related to the scattering
volume. Since the subleading tail ($k^{-4}$) of the large-momentum
distribution cannot fully be captured by the contacts defined by the
adiabatic relations, an extra term resulted from the center-of-mass
motions of the pairs gives rise to an additional divergence in the
kinetic energy of the system, besides those related to the contacts
defined. We show in Tan's energy theorem that if only two-body correlations
are taken into account, all these divergences are reasonably removed,
leading to a finite internal energy of the system. In addition, we
find that all the other universal relations, such as the high-frequency
behavior of the radio-frequency response, short-range behavior of
the pair correlation function, generalized virial theorem, and pressure
relation, remain unaffected by the center-of-mass motions of the pairs,
and are fully governed by the contacts defined by the adiabatic relations.
Our results confirm the feasibility of generalizing the contact theory
for higher-partial-wave scatterings, and could readily be confirmed
in current experiments with ultracold $^{40}$K and $^{6}$Li atoms.
\end{abstract}
\maketitle

\section{Introduction}

In the past decades, ultracold Fermi gases with short-range interactions
have attracted a great deal of interests due to their unique properties
\cite{Bloch2008M,Giorgini2008T}. Especially, near scattering resonances,
where the scattering length $a$ is much larger than all the other
length scales, such systems manifest \emph{universality}: the many-body
properties at long distance are primarily determined by $a$, and
become irrelevant to the specific form of the short-range interatomic
interactions \cite{Braaten2006U}. For strongly interacting two-component
Fermi gases with $s$-wave interactions, a set of universal relations
that follow from the short-range behavior of the simple two-body physics
were derived by Shina Tan, governing the key properties of many-body
systems \cite{Tan2008}. Afterwards, more universal relations were
obtained \cite{Zwerger2011B}. All these relations are characterized
by the only universal quantity named \emph{contact}, and then the
concept of contact becomes significantly important in ultracold atoms
both theoretically and experimentally \cite{Punk2007T,Braaten2008E,Zhang2009U,Schneider2010U,Werner2012G,Hu2010S,Stewart2010V,Kuhnle2010U,Sagi2012M,Hoinka2013P,Valiente2011,Peng2018C}.

However, for higher partial waves, Tan's universal relations should
be amended, since the short-distance behavior of interatomic interactions
cannot simply be characterized by a single scattering parameter. More
microscopic parameters need to be involved besides the scattering
length (or scattering volume, or some quantity like that), such as
the effective range, which may result in non-trivial corrections.
As the simplest case of higher partial wave scatterings, the $p$-wave
many-body systems have attracted both experimental and theoretical
attention \cite{Regal2003T,Zhang2004P,Gunter2005P,Gaebler2007P,Ohashi2005B,Ho2005F,Cheng2005A,Iskin2006E,Idziaszek2006P,Idziaszek2009A,Peng2011,Peng2014}.
Considering the finite-range effect, more contacts are needed when
generalizing the $s$-wave contact theory to the $p$-wave case \cite{Yu2015U,He2016C,Peng2016L,Luciuk2016E}.
It is found that if one tries to define the contacts according to
the adiabatic relations, the subleading tail of the large-momentum
distribution cannot fully be captured, and an extra term appears,
due to the center-of- mass (c.m.) motions of Cooper pairs \cite{Peng2016L}.
This is a general feature of strongly interacting Fermi gases near
$p$-wave resonances. Very recently, the $p$-wave contacts defined
by the adiabatic relations for a two-dimensional (2D) Fermi gas are
discussed in \cite{Zhang2017S}, and even the three-body contact is
introduced in \cite{Zhang2016S} when taking the super Efimov effect
into account. However, the full set of universal relations are still
lack of being justified, such as the energy theorem, the short-distance
behavior of the pair correlation function, and so on.

In this paper, we systematically study the full set of the $p$-wave
universal relations, choosing the 2D spin-polarized Fermi gas as the
model system. We present a derivation of the universal relations following
the route of Tan's original work about the $s$-wave case, in which
only two-body correlations are taken into account \cite{Tan2008}.
Among these universal relations for $p$-wave interactions, the energy
theorem is of particular interest, since it is directly related to
the feasibility of the contact interaction. For $s$-wave interactions,
the kinetic and interaction energies are both ultraviolet divergent
in the zero-range limit, but these divergences cancel with each other
when they add up. The resulting internal energy of the system remains
physically finite, which can be expressed using Tan\textquoteright s
energy theorem, involving only the momentum distribution and the contact.
Here, we show that the internal energy as a functional of the momentum
distribution still exists for a spin-polarized Fermi gas near $p$-wave
resonances in two dimensions, and thereby establish Tan\textquoteright s
energy theorem for $p$-wave interactions. The derivation of the $p$-wave
energy theorem is nontrivial. Unlike the $s$-wave case, the many-body
wave function of a $p$-wave system may not be well normalized in
the zero-range limit. It actually diverges logarithmically as the
interaction range vanishes. Starting from the short-range behavior
of the many-body wave function, we define the $p$-wave contacts according
to the adiabatic relations, and then verify the behavior of the momentum
distribution at large $k$ \cite{Zhang2017S}. We find that both the
leading ($k^{-2}$) and subleading ($k^{-4}$) tails give rise to
the ultraviolet divergence for the kinetic energy. While the subleading
tail cannot fully be described by the contacts defined by the adiabatic
relations, an additional divergence for the kinetic energy arises
due to the c.m. motions of the pairs, besides those related to the
contacts. Here, we demonstrate that all these divergences can reasonably
be removed, leading to a well-defined internal energy of the system.

The high-frequency tail of the radio-frequency (rf) response of the
system is also governed by the contacts, and is experimentally used
as a way to measure the contacts. It links to the momentum distribution
$n\left({\bf k}\right)$ as $\sum_{{\bf k}}n\left({\bf k}\right)\delta\left(\hbar\omega-\hbar^{2}k^{2}/M\right)$,
as the rf frequency $\omega\rightarrow\infty$, a result first derived
by Schneider and Randeria according to the properties of the spectral
function \cite{Schneider2010U}. Here $\hbar$ is the Planck's constant
and $M$ is the atomic mass. At first glance, the c.m. contribution
of the pairs in the subleading $k^{-4}$ tail of the momentum distribution
$n\left({\bf k}\right)$ should be involved in the asymptotic behavior
of the rf response at high frequencies. However, after a rigorous
calculation according to the Fermi's golden rule, we find that the
high-frequency tail of the rf response is determined by $\sum_{{\bf k}}n^{\prime}\left({\bf k}\right)\delta\left(\hbar\omega-\Delta E\right)$,
where $\Delta E$ is the energy difference between the final state
after the rf transition and the initial state, and $n^{\prime}\left({\bf k}\right)$
is not exactly the momentum distribution of the system (see Eq.(\ref{eq:4.11})).
$n^{\prime}\left({\bf k}\right)$ has the same leading behavior as
that of the momentum distribution $n\left({\bf k}\right)$, but different
subleading behavior, in which the c.m. contribution is excluded. After
carefully dealing with this, we finally discover that the high-frequency
tail of the rf response is fully described by the contacts defined
by the adiabatic relations.

In addition, we also obtain the short-distance behavior of the pair
correlation function, which is determined merely by the short-range
behavior of the relative motions of the pairs. Naturally, it is fully
captured by the contacts we defined. Finally, we derive the generalized
virial theorem as well as the pressure relation. These thermodynamic
relations are easily derived by using the adiabatic relations, and
obviously, can fully be described by the contacts defined by the adiabatic
relations. 

This paper is arranged as follows. In Sec.\ref{sec:Adiabatic-relations},
we present the definitions of the $p$-wave contacts, and derive the
specific form of the adiabatic relations for a 2D spin-polarized Fermi
gas. The asymptotic behavior of the momentum distribution at large
momentum is discussed in Sec.\ref{sec:Tail-of-theMomentumDistribution}.
In Sec.\ref{sec:Energy-theorem}, we derive the Tan's energy theorem
for $p$-wave interactions, in which the internal energy of the system
is expressed as a functional of the momentum distribution, and demonstrate
how all the divergences are removed. The high-frequency behavior of
the rf response of the system is studied in Sec.\ref{sec:rf response}
according to the Fermi's golden rule, and in Sec.\ref{sec:Pair-correlation-function},
the short-distance behavior of pair correlation function is obtained.
The general virial theorem is acquired by using the adiabatic relations
as well as the pressure relation in Sec.\ref{sec:Generalized-virial-theorem}.
Finally, our main results are summarized in Sec.\ref{sec:Conclusions}. 

\section{Adiabatic relations\label{sec:Adiabatic-relations}}

Let us consider a strongly interacting spin-polarized Fermi gas with
total particle number $N$ in two dimensions. The interatomic collision
is dominated by the $p$-wave interaction with a short range $\epsilon$,
which is much smaller than all the other length scales of the system.
Then we may deal with the interaction by setting a short-range boundary
condition on many-body wave functions: when any two of fermions, for
example, $i$ and $j$, get close to each other, a many-body wave
function in two dimensions can be written as
\begin{equation}
\Psi_{2D}\left({\bf X},{\bf R},{\bf r}\right)=\sum_{\sigma=\pm}\mathcal{A}_{\sigma}\left({\bf X},{\bf R}\right)\psi_{\sigma}\left({\bf r}\right),\label{eq:1.1}
\end{equation}
 where ${\bf r}={\bf r}_{i}-{\bf r}_{j}$ , ${\bf R}=\left({\bf r}_{i}+{\bf r}_{j}\right)/2$
are, respectively, the relative and c.m. coordinates of the pair $\left(i,j\right)$,
${\bf X}$ includes the degrees of freedom of all the other fermions,
and the index $\sigma=\pm$ denotes two different magnetic components
of the $p$-wave wave function. The function $\mathcal{A}_{\sigma}\left({\bf X},{\bf R}\right)$
is regular and $\psi_{\sigma}\left({\bf r}\right)$ is the two-body
wave function describing the relative motion of the pair, which should
take the form (unnormalized)
\begin{equation}
\psi_{\sigma}\left({\bf r}\right)=\frac{\pi}{2}q\left[J_{1}\left(qr\right)\cot\delta_{\sigma}-N_{1}\left(qr\right)\right]\Omega_{1}^{(\sigma)}\left(\varphi\right)\label{eq:1.2}
\end{equation}
outside the range of the interatomic interaction, where $J_{\nu}\left(\cdot\right)$,
$N_{\nu}\left(\cdot\right)$ are the Bessel functions of the first
and second kinds, $q$ is the relative wave number of the pair, $\Omega_{m}^{(\sigma)}\left(\varphi\right)\equiv e^{i\sigma m\varphi}/\sqrt{2\pi}$
is the angular function with respect to the azimuthal angle $\varphi$
of the vector ${\bf r}$. We note that the \emph{regularity} of the
function $\mathcal{A}_{\sigma}\left({\bf X},{\bf R}\right)$ implies
that no more pairs except the fermions $i$ and $j$ can interact
with each other, because of such a short-range interaction.

The interactions between spin-polarized fermions can be tuned using
$p$-wave Feshbach resonances experimentally, and Tan's adiabatic
relations state how the total energy of the system accordingly changes
when the interatomic interaction is adiabatically adjusted. To derive
the adiabatic relations, we consider two many-body wave functions
$\Psi_{2D}$ and $\Psi_{2D}^{\prime}$ corresponding to different
interaction strengths, and they should satisfy the Schr\"{o}dinger
equations with different energies
\begin{eqnarray}
\sum_{i=1}^{N}\left[-\frac{\hbar^{2}}{2M}\nabla_{i}^{2}+U\left({\bf r}_{i}\right)\right]\Psi_{2D} & = & E\Psi_{2D},\label{eq:1.3}\\
\sum_{i=1}^{N}\left[-\frac{\hbar^{2}}{2M}\nabla_{i}^{2}+U\left({\bf r}_{i}\right)\right]\Psi_{2D}^{\prime} & = & E^{\prime}\Psi_{2D}^{\prime},\label{eq:1.4}
\end{eqnarray}
 if there is no pair of fermions within the range of the interaction.
Here, $M$ is the atomic mass and $U\left({\bf r}_{i}\right)$ is
the external potential experienced by the $i$-th fermion. Then it
follows from Eqs.(\ref{eq:1.3}) and (\ref{eq:1.4}) that \cite{Peng2016L}
\begin{multline}
\left(E-E^{\prime}\right)\int_{\mathcal{S}_{\epsilon}}\prod_{i=1}^{N}d{\bf r}_{i}\Psi_{2D}^{\prime*}\Psi_{2D}=\\
-\frac{\hbar^{2}}{M}\mathcal{N}\oint_{r=\epsilon}\left(\Psi_{2D}^{\prime*}\nabla_{{\bf r}}\Psi_{2D}-\Psi_{2D}\nabla_{{\bf r}}\Psi_{2D}^{\prime*}\right)\cdot\hat{{\bf n}}dl,\label{eq:1.5}
\end{multline}
 where $\mathcal{N}=N\left(N-1\right)/2$ is the number of all the
possible ways to pair atoms, the domain $\mathcal{S}_{\epsilon}$
is the set of all configurations $\left({\bf r}_{i},{\bf r}_{j}\right)$,
in which $r=\left|{\bf r}_{i}-{\bf r}_{j}\right|>\epsilon$, $l$
is the boundary of $\mathcal{S}_{\epsilon}$ that the distance between
the two fermions in the pair $\left(i,j\right)$ is $\epsilon$, and
$\hat{{\bf n}}$ is the direction normal to $l$, but is opposite
to the radial direction. Expanding the many-body wave function (\ref{eq:1.1})
at small $r$, we obtain
\begin{multline}
\Psi_{2D}\left({\bf X},{\bf R},{\bf r}\right)\approx\sum_{\sigma}\mathcal{A}_{\sigma}\left({\bf X},{\bf R}\right)\left[\frac{1}{r}-\frac{q^{2}}{2}r\ln\frac{r}{2b_{\sigma}}\right.\\
\left.-\left(\frac{\pi}{4a_{\sigma}}+\frac{2\gamma-1}{4}q^{2}\right)r+\mathcal{O}\left(r^{3}\right)\right]\Omega_{1}^{(\sigma)}\left(\varphi\right),\label{eq:1.6}
\end{multline}
 where $\gamma$ is the Euler's constant, and we have used the effective-range
expansion of the $p$-wave scattering phase shift for 2D systems \cite{Zhang2018H},
i.e.,
\begin{equation}
\cot\delta_{\sigma}=-\frac{1}{a_{\sigma}q^{2}}+\frac{2}{\pi}\ln\left(qb_{\sigma}\right),\label{eq:1.7}
\end{equation}
 and $a_{\sigma},b_{\sigma}$ are the scattering area and effective
range, respectively, with the dimensions of length$^{2}$ and length$^{1}$
. We should note that the pair relative wave number $q$ is generally
dependent on ${\bf X}$ as well as ${\bf R}$, due to the external
confinement $U$ and the interatomic interactions, and can formally
be written as \cite{Peng2016L,Werner2012G}
\begin{equation}
\frac{\hbar^{2}q^{2}}{M}=E-\frac{1}{\mathcal{A}_{\sigma}\left({\bf X},{\bf R}\right)}\left[T\left({\bf X},{\bf R}\right)+U\left({\bf X},{\bf R}\right)\right]\mathcal{A}_{\sigma}\left({\bf X},{\bf R}\right),\label{eq:1.8}
\end{equation}
and $T\left({\bf X},{\bf R}\right)$ and $U\left({\bf X},{\bf R}\right)$
are respectively the kinetic and external potential operators including
the c.m. motion of the pair $\left(i,j\right)$ and those of the rest
of the fermions. Inserting the asymptotic form of the many-body wave
function (\ref{eq:1.6}) into Eq.(\ref{eq:1.5}), and letting $E^{\prime}\rightarrow E$,
$a_{\sigma}^{\prime}\rightarrow a_{\sigma}$, $b_{\sigma}^{\prime}\rightarrow b_{\sigma}$,
we easily obtain
\begin{multline}
\delta E\cdot\int_{\mathcal{S}_{\epsilon}}\prod_{i=1}^{N}d{\bf r}_{i}\left|\Psi_{2D}\right|^{2}=\sum_{\sigma}\left[-\frac{\pi\hbar^{2}}{2M}\mathcal{N}\mathcal{I}_{a}^{(\sigma)}\cdot\delta a_{\sigma}^{-1}\right.\\
\left.+\mathcal{N}\mathcal{E}_{\sigma}\cdot\delta\ln b_{\sigma}+\mathcal{N}\mathcal{I}_{a}^{(\sigma)}\left(\ln\frac{2b_{\sigma}}{\epsilon}-\gamma\right)\cdot\delta E\right],\label{eq:1.9}
\end{multline}
 where
\begin{eqnarray}
\mathcal{I}_{a}^{(\sigma)} & \equiv & \int d{\bf X}d{\bf R}\left|\mathcal{A}_{\sigma}\left({\bf X},{\bf R}\right)\right|^{2},\label{eq:1.10}\\
\mathcal{E}_{\sigma} & \equiv & \int d{\bf X}d{\bf R}\mathcal{A}_{\sigma}^{*}\left(E-T-U\right)\mathcal{A}_{\sigma}.\label{eq:1.11}
\end{eqnarray}
 Using the normalization of the wave function (see Appendix \ref{sec:TheNormalization})
\begin{equation}
\int_{\mathcal{S}_{\epsilon}}\prod_{i=1}^{N}d{\bf r}_{i}\left|\Psi_{2D}\right|^{2}=1+\mathcal{N}\sum_{\sigma}\mathcal{I}_{a}^{(\sigma)}\left(\ln\frac{2b_{\sigma}}{\epsilon}-\gamma\right),\label{eq:1.12}
\end{equation}
 Eq.(\ref{eq:1.9}) can further be simplified as
\begin{equation}
\delta E=\sum_{\sigma}\left[-\frac{\pi\hbar^{2}}{2M}\mathcal{N}\mathcal{I}_{a}^{(\sigma)}\cdot\delta a_{\sigma}^{-1}+\mathcal{N}\mathcal{E}_{\sigma}\cdot\delta\ln b_{\sigma}\right],\label{eq:1.13}
\end{equation}
 which yields
\begin{eqnarray}
\frac{\partial E}{\partial a_{\sigma}^{-1}} & = & -\frac{\pi\hbar^{2}}{2M}\mathcal{N}\mathcal{I}_{a}^{(\sigma)},\label{eq:1.14}\\
\frac{\partial E}{\partial\ln b_{\sigma}} & = & \mathcal{N}\mathcal{E}_{\sigma}.\label{eq:1.15}
\end{eqnarray}

\section{Tail of the momentum distribution at large $k$ and contacts\label{sec:Tail-of-theMomentumDistribution}}

In this section, we are going to study the asymptotic behavior of
the large momentum distribution for a spin-polarized Fermi gas. The
momentum distribution of the $i$-th fermion is defined as
\begin{equation}
n_{i}\left({\bf k}\right)\equiv\int\prod_{t\neq i}d{\bf r}_{t}\left|\tilde{\Psi}_{i}\left({\bf k}\right)\right|^{2},\label{eq:2.1}
\end{equation}
where $\tilde{\Psi}_{i}\left({\bf k}\right)\equiv\int d{\bf r}_{i}\Psi_{2D}e^{-i{\bf k}\cdot{\bf r}_{i}}$,
and then the total momentum distribution is $n\left({\bf k}\right)=\sum_{i=1}^{N}n_{i}\left({\bf k}\right)$.
When the pair $\left(i,j\right)$ get close but still outside the
interaction range, i.e., $r\left(\sim0^{+}\right)>\epsilon$, while
all the other fermions are far away, we may again expand the many-body
wave function $\Psi_{2D}$ (\ref{eq:1.1}) at ${\bf r}\approx0$,
and rewrite it as the following ansatz
\begin{multline}
\Psi_{2D}\left({\bf X},{\bf R},{\bf r}\right)=\sum_{\sigma}\left[\frac{\mathcal{A}_{\sigma}\left({\bf X},{\bf R}\right)}{r}+\mathcal{B}_{\sigma}\left({\bf X},{\bf R}\right)r\ln r\right.\\
\left.+\mathcal{C}_{\sigma}\left({\bf X},{\bf R}\right)r\right]\Omega_{1}^{(\sigma)}\left(\varphi\right)+{\bf r}\cdot{\bf L}\left({\bf X},{\bf R}\right)+\mathcal{O}\left(r^{3}\right),\label{eq:2.2}
\end{multline}
 where $\mathcal{A}_{\sigma}$, $\mathcal{B}_{\sigma}$, $\mathcal{C}_{\sigma}$
and ${\bf L}$ are all regular functions, and the term ${\bf r}\cdot{\bf L}\left({\bf X},{\bf R}\right)$
represents the coupling between the relative and c.m. motions of the
pair $\left(i,j\right)$, resulted from the external confinement.
Comparing Eqs.(\ref{eq:1.6}) and (\ref{eq:2.2}) at small $r$, we
find
\begin{eqnarray}
\mathcal{B}_{\sigma} & = & -\frac{q^{2}}{2}\mathcal{A}_{\sigma},\label{eq:2.3}\\
\mathcal{C}_{\sigma} & = & \left(\frac{1-2\gamma}{4}q^{2}-\frac{\pi}{4a_{\sigma}}+\frac{q^{2}}{2}\ln2b_{\sigma}\right)\mathcal{A}_{\sigma}\nonumber \\
 & = & -\left(\frac{1-2\gamma}{2}+\ln2b_{\sigma}\right)\mathcal{B}_{\sigma}-\frac{\pi}{4a_{\sigma}}\mathcal{A}_{\sigma},\label{eq:2.4}
\end{eqnarray}
 and $\mathcal{E}_{\sigma}$ defined in Eq.(\ref{eq:1.11}) can alternatively
be rewritten as
\begin{equation}
\mathcal{E}_{\sigma}=-\frac{2\hbar^{2}}{M}\int d{\bf X}d{\bf R}\mathcal{A}_{\sigma}^{*}\left({\bf X},{\bf R}\right)\mathcal{B}_{\sigma}\left({\bf X},{\bf R}\right)\equiv-\frac{2\hbar^{2}}{M}\mathcal{I}_{b}^{(\sigma)},\label{eq:2.5}
\end{equation}
 where
\begin{equation}
\mathcal{I}_{b}^{(\sigma)}\equiv\int d{\bf X}d{\bf R}\mathcal{A}_{\sigma}^{*}\left({\bf X},{\bf R}\right)\mathcal{B}_{\sigma}\left({\bf X},{\bf R}\right)\label{eq:2.6}
\end{equation}
 is obviously real. The asymptotic behavior of the momentum distribution
at large ${\bf k}$ but still smaller than $\epsilon^{-1}$ is determined
by that of the wave function at short distance, then we have
\begin{equation}
\tilde{\Psi}_{i}\left({\bf k}\right)\underset{k\rightarrow\infty}{\approx}\sum_{j\neq i}e^{-i{\bf k}\cdot{\bf r}_{j}}\int d{\bf r}\Psi_{2D}\left({\bf X},{\bf r}_{j}+\frac{{\bf r}}{2},{\bf r}\right)e^{-i{\bf k}\cdot{\bf r}}.\label{eq:2.7}
\end{equation}
 With the help of the plane-wave expansion 
\begin{equation}
e^{i{\bf k}\cdot{\bf r}}=\sqrt{2\pi}\sum_{m=0}^{\infty}\sum_{\sigma=\pm}\eta_{m}i^{m}J_{m}\left(kr\right)e^{-i\sigma m\varphi_{{\bf k}}}\Omega_{m}^{(\sigma)}\left(\varphi\right),\label{eq:2.8}
\end{equation}
 where $\eta_{m}=1/2$ for $m=0$, and $\eta_{m}=1$ for $m\ge1$
, and $\varphi_{{\bf k}}$ is the azimuthal angle of ${\bf k}$, we
find
\begin{align}
 & \int d{\bf r}\frac{\mathcal{A}_{\sigma}\left({\bf X},{\bf r}_{j}+{\bf r}/2\right)}{r}\Omega_{1}^{(\sigma)}\left(\varphi\right)e^{-i{\bf k}\cdot{\bf r}}\nonumber \\
= & -i\sqrt{2\pi}\mathcal{A}_{\sigma}\left({\bf X},{\bf r}_{j}\right)\frac{e^{i\sigma\varphi_{{\bf k}}}}{k}+\sqrt{\frac{\pi}{2}}\alpha_{\sigma}\left({\bf X},{\bf r}_{j},\hat{{\bf k}}\right)\frac{1}{k^{2}}\nonumber \\
 & +i\frac{\sqrt{2\pi}}{8}\beta_{\sigma}\left({\bf X},{\bf r}_{j},\hat{{\bf k}}\right)\frac{1}{k^{3}}+\mathcal{O}\left(k^{-4}\right),\label{eq:2.9}
\end{align}
 where
\begin{eqnarray}
\alpha_{\sigma}\left({\bf X},{\bf r}_{j},\hat{{\bf k}}\right) & \equiv & k^{2}\nabla_{{\bf r}_{j}}\mathcal{A}_{\sigma}\cdot\nabla_{{\bf k}}\frac{e^{i\sigma\varphi_{{\bf k}}}}{k},\label{eq:2.10}\\
\beta_{\sigma}\left({\bf X},{\bf r}_{j},\hat{{\bf k}}\right) & \equiv & k^{3}\left(\nabla_{{\bf r}_{j}}\cdot\nabla_{{\bf k}}\right)\left[\nabla_{{\bf r}_{j}}\mathcal{A}_{\sigma}\cdot\nabla_{{\bf k}}\frac{e^{i\sigma\varphi_{{\bf k}}}}{k}\right]\label{eq:2.11}
\end{eqnarray}
only depend on the direction of ${\bf k}$,
\begin{multline}
\int d{\bf r}\mathcal{B}_{\sigma}\left({\bf X},{\bf r}_{j}+\frac{{\bf r}}{2}\right)\left(r\ln r\right)\Omega_{1}^{(\sigma)}\left(\varphi\right)e^{-i{\bf k}\cdot{\bf r}}\\
=i2\sqrt{2\pi}\mathcal{B}_{\sigma}\left({\bf X},{\bf r}_{j}\right)\frac{e^{i\sigma\varphi_{{\bf k}}}}{k^{3}}+\mathcal{O}\left(k^{-4}\right),\label{eq:2.12}
\end{multline}
and
\begin{equation}
\int d{\bf r}\mathcal{C}_{\sigma}\left({\bf X},{\bf r}_{j}+\frac{{\bf r}}{2}\right)r\Omega_{1}^{(\sigma)}\left(\varphi\right)e^{-i{\bf k}\cdot{\bf r}}=0.\label{eq:2.13}
\end{equation}
 In addition, it is obvious that the coupling term ${\bf r}\cdot{\bf L}\left({\bf X},{\bf r}_{j}+{\bf r}/2\right)$
contributes nothing to the tail of the momentum distribution at large
$k$. Therefore, inserting Eqs.(\ref{eq:2.9}), (\ref{eq:2.12}),
and (\ref{eq:2.13}) into Eq.(\ref{eq:2.7}), and then into Eq.(\ref{eq:2.1}),
we find the total momentum distribution $n\left({\bf k}\right)$ at
large ${\bf k}$ takes the form\begin{widetext}
\begin{multline}
n\left({\bf k}\right)\approx\mathcal{N}\int d{\bf X}d{\bf R}\sum_{\sigma\sigma^{\prime}}\mathcal{A}_{\sigma^{\prime}}^{*}\mathcal{A}_{\sigma}e^{i\left(\sigma-\sigma^{\prime}\right)\varphi_{{\bf k}}}\cdot\frac{4\pi}{k^{2}}+\text{{\bf Im}}\mathcal{N}\int d{\bf X}d{\bf R}\sum_{\sigma\sigma^{\prime}}\mathcal{A}_{\sigma}e^{i\sigma\varphi_{{\bf k}}}\alpha_{\sigma^{\prime}}^{*}\cdot\frac{4\pi}{k^{3}}\\
+\left[-16\pi\text{{\bf Re}}\mathcal{N}\int d{\bf X}d{\bf R}\sum_{\sigma\sigma^{\prime}}\mathcal{A}_{\sigma^{\prime}}^{*}\mathcal{B}_{\sigma}e^{i\left(\sigma-\sigma^{\prime}\right)\varphi_{{\bf k}}}+\pi\text{{\bf Re}}\mathcal{N}\int d{\bf X}d{\bf R}\sum_{\sigma\sigma^{\prime}}\left(\alpha_{\sigma^{\prime}}^{*}\alpha_{\sigma}-\mathcal{A}_{\sigma}e^{i\sigma\varphi_{{\bf k}}}\beta_{\sigma^{\prime}}^{*}\right)\right]\frac{1}{k^{4}}+\mathcal{O}\left(k^{-5}\right),\label{eq:2.14}
\end{multline}
\end{widetext}where we have rewritten the integral variable ${\bf r}_{j}$
as ${\bf R}$, and we have also omitted the arguments of the functions
$\mathcal{A}$, $\mathcal{B}$, $\alpha_{\sigma}$, and $\beta_{\sigma}$
to simplify the expression. If we are only interested in the dependence
of the momentum distribution on the amplitude of ${\bf k}$, we may
integrate over the direction of ${\bf k}$, and we find all the odd-order
terms of $k^{-1}$ vanish. We obtain (see Appendix \ref{sec:MomentumDistribution})
\begin{equation}
n\left(k\right)\approx\frac{\sum_{\sigma}\mathcal{C}_{a}^{(\sigma)}}{k^{2}}+\sum_{\sigma}\left(\mathcal{C}_{b}^{(\sigma)}+\mathcal{Q}_{cm}^{(\sigma)}\right)\frac{1}{k^{4}}+\mathcal{O}\left(k^{-6}\right),\label{eq:2.15}
\end{equation}
where the contacts $\mathcal{C}_{a}^{(\sigma)}$ and $\mathcal{C}_{b}^{(\sigma)}$
are defined as
\begin{eqnarray}
\mathcal{C}_{a}^{(\sigma)} & \equiv & 8\pi^{2}\mathcal{N}\mathcal{I}_{a}^{(\sigma)},\label{eq:2.16}\\
\mathcal{C}_{b}^{(\sigma)} & \equiv & -32\pi^{2}\mathcal{N}\mathcal{I}_{b}^{(\sigma)},\label{eq:2.17}
\end{eqnarray}
and
\begin{equation}
\mathcal{Q}_{cm}^{(\sigma)}\equiv2\pi^{2}\mathcal{N}\int d{\bf X}d{\bf R}\left(\nabla_{{\bf R}}\mathcal{A}_{\sigma}^{*}\cdot\nabla_{{\bf R}}\mathcal{A}\right).
\end{equation}
Therefore, the adiabatic relations (\ref{eq:1.14}) and (\ref{eq:1.15})
can alternatively be written as
\begin{eqnarray}
\frac{\partial E}{\partial a_{\sigma}^{-1}} & = & -\frac{\hbar^{2}\mathcal{C}_{a}^{(\sigma)}}{16\pi M},\label{eq:2.18}\\
\frac{\partial E}{\partial\ln b_{\sigma}} & = & \frac{\hbar^{2}\mathcal{C}_{b}^{(\sigma)}}{16\pi^{2}M}.\label{eq:2.19}
\end{eqnarray}
 We find, similarly as the situation in three dimensions \cite{Peng2016L},
the leading-order term of $k^{-2}$ can fully be described by the
contact $\mathcal{C}_{a}^{(\sigma)}$, while there is an extra term
appearing in the subleading-order term of $k^{-4}$, i.e., $\mathcal{Q}_{cm}^{(\sigma)}$,
in addition to the contact $\mathcal{C}_{b}^{(\sigma)}$, which is
resulted from the c.m. motions of the pairs. We can expect that this
additional term should result in significant amendments to the other
universal relations.

\section{Energy theorem \label{sec:Energy-theorem}}

Because of the short-range $p$-wave interatomic interactions, the
momentum distribution generally decays like $k^{-2}$ at large ${\bf k}$,
and subsequently the kinetic energy of the system diverges. Unlike
that of the $s$-wave interaction, the subleading-order term of $k^{-4}$
in the large momentum distribution should also result in an additional
divergence of the kinetic energy. In addition, such divergent behavior
in the subleading-order term can fully be captured only when both
the contact $\mathcal{C}_{b}^{(\sigma)}$ defined from adiabatic relation
(\ref{eq:2.19}) and the extra term $\mathcal{Q}_{cm}^{(\sigma)}$
resulted from the c.m. motions of the pairs are considered. In this
section, we show that all these divergences can be removed, leading
to a convergent total internal energy and the $p$-wave Tan\textquoteright s
energy theorem.

In the follows, we take only two-body correlations into account, which
should be reasonable at two-body resonances, and all higher-order
correlations can be neglected. Therefore, in order to avoid the complication
of the notations, we first demonstrate the derivation of the energy
theorem according to a two-body picture, and then present the general
energy theorem for a many-body system. Because only the internal energy
of the system is considered, we are going to omit the external confinement,
which is trivial to the energy theorem. The Schr\"{o}dinger equation
of two fermions takes the form
\begin{equation}
E\Psi_{2D}=\left[\sum_{i=1}^{2}\left(-\frac{\hbar^{2}}{2M}\nabla_{i}^{2}\right)+V\left({\bf r}_{1}-{\bf r}_{2}\right)\right]\Psi_{2D},\label{eq:3.1}
\end{equation}
 where $V\left({\bf r}_{1}-{\bf r}_{2}\right)$ is the interatomic
interaction with a short range $\epsilon$, out of which we may assume
$V=0$. Multiplying $\Psi_{2D}^{*}$ and integrating on both sides
of Eq.(\ref{eq:3.1}) over the domain $s_{\epsilon}$, in which $r=\left|{\bf r}_{1}-{\bf r}_{2}\right|>\epsilon$,
we obtain
\begin{multline}
E\int_{s_{\epsilon}}d{\bf r}_{1}d{\bf r}_{2}\left|\Psi_{2D}\right|^{2}=\\
\int_{s_{\epsilon}}d{\bf r}_{1}d{\bf r}_{2}\Psi_{2D}^{*}\sum_{i=1}^{2}\left(-\frac{\hbar^{2}}{2M}\nabla_{i}^{2}\right)\Psi_{2D}.\label{eq:3.2}
\end{multline}
 On the left-hand side (LHS) of Eq.(\ref{eq:3.2}), we already obtain
\begin{equation}
\int_{s_{\epsilon}}d{\bf r}_{1}d{\bf r}_{2}\left|\Psi_{2D}\right|^{2}=1+\sum_{\sigma}\mathcal{I}_{a}^{(\sigma)}\left(\ln\frac{2b_{\sigma}}{\epsilon}-\gamma\right)\label{eq:3.3}
\end{equation}
in Appendix \ref{sec:TheNormalization}. Let us concentrate on the
right-hand side (RHS), which may be rewritten as
\begin{equation}
RHS=I_{whole}^{(2)}-I_{\bar{s}_{\epsilon}}^{(2)},\label{eq:3.4}
\end{equation}
where
\begin{eqnarray}
I_{whole}^{(2)} & \equiv & \int d{\bf r}_{1}d{\bf r}_{2}\Psi_{2D}^{*}\sum_{i=1}^{2}\left(-\frac{\hbar^{2}}{2M}\nabla_{i}^{2}\right)\Psi_{2D},\label{eq:3.5}\\
I_{\bar{s}_{\epsilon}}^{(2)} & \equiv & \int_{\bar{s}_{\epsilon}}d{\bf r}_{1}d{\bf r}_{2}\Psi_{2D}^{*}\sum_{i=1}^{2}\left(-\frac{\hbar^{2}}{2M}\nabla_{i}^{2}\right)\Psi_{2D}.\label{eq:3.6}
\end{eqnarray}
 Here, $\bar{s}_{\epsilon}$ is the complementary set of $s_{\epsilon}$,
in which $r<\epsilon$. If we write the two-body wave function $\Psi_{2D}$
in the momentum space, i.e.,
\begin{equation}
\Psi_{2D}=\sum_{{\bf k}_{1}}\sum_{{\bf k}_{2}}\Phi_{2D}\left({\bf k}_{1},{\bf k}_{2}\right)e^{i{\bf k}_{1}\cdot{\bf r}_{1}}e^{i{\bf k}_{2}\cdot{\bf r}_{2}},\label{eq:3.7}
\end{equation}
where
\begin{equation}
\Phi_{2D}\left({\bf k}_{1},{\bf k}_{2}\right)\equiv\int d{\bf r}_{1}d{\bf r}_{2}\Psi_{2D}e^{-i{\bf k}_{1}\cdot{\bf r}_{1}}e^{-i{\bf k}_{2}\cdot{\bf r}_{2}},\label{eq:3.8}
\end{equation}
 $I_{whole}^{(2)}$ becomes
\begin{eqnarray}
I_{whole}^{(2)} & = & \sum_{{\bf k}_{1}}\sum_{{\bf k}_{2}}\sum_{i=1}^{2}\frac{\hbar^{2}k_{i}^{2}}{2M}\left|\Phi_{2D}\right|^{2}\nonumber \\
 & = & \sum_{{\bf k}}\frac{\hbar^{2}k^{2}}{2M}n\left({\bf k}\right),\label{eq:3.9}
\end{eqnarray}
 where $n\left({\bf k}\right)$ is the total momentum distribution
of two fermions. If we extend the asymptotic form of $\Psi_{2D}$
(\ref{eq:1.1}) to the region even inside the interaction range, the
momentum distribution $n\left(k\right)$ decays like $k^{-2}$ and
$k^{-4}$ as $k\rightarrow\infty$, respectively, as shown in Eq.(\ref{eq:2.15}),
and then $I_{whole}^{(2)}$ becomes divergent. However, we will see
such divergence is exactly removed by $I_{\bar{s}_{\epsilon}}^{(2)}$,
and the RHS of Eq.(\ref{eq:3.2}), i.e., Eq.(\ref{eq:3.4}), converges. 

In the follows, let us focus on the integral $I_{\bar{s}_{\epsilon}}$.
Inserting Eq.(\ref{eq:1.1}) into Eq.(\ref{eq:3.6}), and rewriting
the integral in the c.m. frame of two fermions, we obtain
\begin{multline}
I_{\bar{s}_{\epsilon}}^{(2)}=\sum_{\sigma\sigma^{\prime}}\left[\int d{\bf R}\mathcal{A}_{\sigma^{\prime}}^{*}\mathcal{A}_{\sigma}\cdot{\bf I}_{1}^{(\sigma\sigma^{\prime})}\right.\\
\left.+\int d{\bf R}\mathcal{A}_{\sigma^{\prime}}^{*}\left(-\frac{\hbar^{2}}{4M}\nabla_{{\bf R}}^{2}\right)\mathcal{A}_{\sigma}\cdot{\bf I}_{2}^{(\sigma\sigma^{\prime})}\right],\label{eq:3.10}
\end{multline}
 where 
\begin{eqnarray}
{\bf I}_{1}^{(\sigma\sigma^{\prime})} & \equiv & \int_{r<\epsilon}d{\bf r}\psi_{\sigma^{\prime}}^{*}\left({\bf r}\right)\left(-\frac{\hbar^{2}}{M}\nabla_{{\bf r}}^{2}\right)\psi_{\sigma}\left({\bf r}\right),\label{eq:3.11}\\
{\bf I}_{2}^{(\sigma\sigma^{\prime})} & \equiv & \int_{r<\epsilon}d{\bf r}\psi_{\sigma^{\prime}}^{*}\left({\bf r}\right)\psi_{\sigma}\left({\bf r}\right),\label{eq:3.12}
\end{eqnarray}
 and we should note that the variable ${\bf X}$ in the function $\mathcal{A}$
drops out automatically for a two-body system. Let us calculate ${\bf I}_{1}^{(\sigma\sigma^{\prime})}$
first, and keep in mind that $\psi_{\sigma}\left({\bf r}\right)$
takes the form of Eq.(\ref{eq:1.2}), which is the linear combination
of $J_{1}\left(qr\right)\Omega_{1}^{(\sigma)}\left(\varphi\right)$
and $N_{1}\left(qr\right)\Omega_{1}^{(\sigma)}\left(\varphi\right)$,
respectively, the regular and irregular solutions of the $p$-wave
Schr\"{o}dinger equation. Therefore, we have
\begin{equation}
\nabla_{{\bf r}}^{2}\left[J_{1}\left(qr\right)\Omega_{1}^{(\sigma)}\left(\varphi\right)\right]=-q^{2}J_{1}\left(qr\right)\Omega_{1}^{(\sigma)}\left(\varphi\right),\label{eq:3.13}
\end{equation}
and then we easily find
\begin{multline}
\int_{r<\epsilon}d{\bf r}\psi_{\sigma^{\prime}}^{*}\left({\bf r}\right)\left(-\frac{\hbar^{2}}{M}\nabla_{{\bf r}}^{2}\right)\left[J_{1}\left(qr\right)\Omega_{1}^{(\sigma)}\left(\varphi\right)\right]\\
=\delta_{\sigma\sigma^{\prime}}\left(\frac{\hbar^{2}q}{4M}\right)\left[\left(\epsilon q\right)^{2}+O\left(\epsilon q\right)^{4}\right],\label{eq:3.14}
\end{multline}
 which vanishes in the low-energy limit, i.e., $\epsilon q\rightarrow0$,
and it yields
\begin{equation}
{\bf I}_{1}^{(\sigma\sigma^{\prime})}=\frac{\pi\hbar^{2}q}{2M}\int_{r<\epsilon}d{\bf r}\psi_{\sigma^{\prime}}^{*}\left({\bf r}\right)\nabla_{{\bf r}}^{2}\left[N_{1}\left(qr\right)\Omega_{1}^{(\sigma)}\left(\varphi\right)\right].\label{eq:3.15}
\end{equation}
 As to the irregular solution $N_{1}\left(qr\right)\Omega_{1}^{(\sigma)}\left(\varphi\right)$,
we have (see Appendix \ref{sec:NormOfBesselFunction})
\begin{equation}
\nabla_{{\bf r}}^{2}\left[N_{1}\left(qr\right)\Omega_{1}^{(\sigma)}\left(\varphi\right)\right]=\left[\frac{4\delta\left(r\right)}{\pi qr^{2}}-q^{2}N_{1}\left(qr\right)\right]\Omega_{1}^{(\sigma)}\left(\varphi\right),\label{eq:3.16}
\end{equation}
 and then
\begin{multline}
{\bf I}_{1}^{(\sigma\sigma^{\prime})}=\delta_{\sigma\sigma^{\prime}}\frac{\hbar^{2}}{M}\left(\frac{\pi}{2}q^{2}\cot\delta_{\sigma}\right)\\
-\frac{\pi^{2}\hbar^{2}q^{2}}{4M}\int_{r<\epsilon}d{\bf r}N_{1}\left(qr\right)\Omega_{1}^{(\sigma^{\prime})*}\left(\varphi\right)\nabla_{{\bf r}}^{2}\left[N_{1}\left(qr\right)\Omega_{1}^{(\sigma)}\left(\varphi\right)\right].\label{eq:3.17}
\end{multline}
Apparently, the last term of Eq.(\ref{eq:3.17}) is divergent, since
the Bessel function $N_{1}\left(qr\right)$ behaves as
\begin{equation}
N_{1}\left(qr\right)=-\frac{2}{\pi qr}+\frac{qr}{\pi}\left(\ln\frac{qr}{2}+\gamma-\frac{1}{2}\right)+O\left(qr\right)^{3}\label{eq:3.18}
\end{equation}
 at $qr\sim0$. The crucial point of the energy theorem is to discuss
such divergence alternatively in the momentum space. After straightforward
algebra, we show in the Appendix \ref{sec:Calculation-details-of}
that\begin{widetext}
\begin{multline}
\frac{\pi^{2}\hbar^{2}q^{2}}{4M}\int_{r<\epsilon}d{\bf r}N_{1}\left(qr\right)\Omega_{1}^{(\sigma^{\prime})*}\left(\varphi\right)\nabla_{{\bf r}}^{2}\left[N_{1}\left(qr\right)\Omega_{1}^{(\sigma)}\left(\varphi\right)\right]\\
=\delta_{\sigma\sigma^{\prime}}\lim_{\Lambda\rightarrow\infty}\left[-\frac{\hbar^{2}\Lambda^{2}}{2M}-\frac{\hbar^{2}q^{2}}{M}\ln\frac{\Lambda}{q}-\frac{\hbar^{2}q^{2}}{M}\left(\gamma+\ln\frac{\epsilon\Lambda}{2}\right)-\int_{\Lambda}^{\infty}\frac{kdk}{\left(2\pi\right)^{2}}\frac{\hbar^{2}k^{2}}{2M}\left(\frac{8\pi^{2}}{k^{2}}+\frac{16\pi^{2}q^{2}}{k^{4}}\right)\right],\label{eq:3.19}
\end{multline}
and then it yields
\begin{equation}
{\bf I}_{1}^{(\sigma\sigma^{\prime})}=\delta_{\sigma\sigma^{\prime}}\lim_{\Lambda\rightarrow\infty}\left[\frac{\hbar^{2}}{2M}\left(\Lambda^{2}-\frac{\pi}{a_{\sigma}}\right)+\frac{\hbar^{2}q^{2}}{M}\ln\left(\Lambda b_{\sigma}\right)+\frac{\hbar^{2}q^{2}}{M}\left(\gamma+\ln\frac{\epsilon\Lambda}{2}\right)+\int_{\Lambda}^{\infty}\frac{kdk}{\left(2\pi\right)^{2}}\frac{\hbar^{2}k^{2}}{2M}\left(\frac{8\pi^{2}}{k^{2}}+\frac{16\pi^{2}q^{2}}{k^{4}}\right)\right],\label{eq:3.20}
\end{equation}
 \end{widetext}where we have used the effective expansion (\ref{eq:1.7}).
In the expression of Eq.(\ref{eq:3.20}), we exactly separate the
divergent part of ${\bf I}_{1}^{(\sigma\sigma^{\prime})}$ in the
momentum space as appearing in the last integral.

As to the integral ${\bf I}_{2}^{(\sigma\sigma^{\prime})}$, i.e.,
Eq.(\ref{eq:3.12}), we easily find it is also divergent, since the
wave function $\psi_{\sigma}\left({\bf r}\right)$ behaves as $r^{-1}$
at $r\sim0$. Such divergence can also be separated in the momentum
space (see Appendix \ref{sec:Calculation-details-of}), and yields
\begin{equation}
{\bf I}_{2}^{(\sigma\sigma^{\prime})}=\delta_{\sigma\sigma^{\prime}}\lim_{\Lambda\rightarrow\infty}\left[\gamma+\ln\frac{\epsilon\Lambda}{2}+4\pi^{2}\int_{\Lambda}^{\infty}\frac{kdk}{\left(2\pi\right)^{2}}\frac{1}{k^{2}}\right].\label{eq:3.21}
\end{equation}
 Combining Eqs.(\ref{eq:3.10}) (\ref{eq:3.20}) and (\ref{eq:3.21}),
we obtain\begin{widetext}
\begin{multline}
I_{\bar{\epsilon}}^{(2)}=\lim_{\Lambda\rightarrow\infty}\left\{ \sum_{\sigma}\frac{\hbar^{2}c_{a}^{(\sigma)}}{16\pi^{2}M}\left(\Lambda^{2}-\frac{\pi}{a_{\sigma}}\right)+\frac{\hbar^{2}}{8\pi^{2}M}\sum_{\sigma}\left[\frac{c_{b}^{(\sigma)}}{2}\ln\left(\Lambda b_{\sigma}\right)+\left(\frac{c_{b}^{(\sigma)}}{2}+Q_{cm}^{(\sigma)}\right)\left(\ln\frac{\epsilon\Lambda}{2}+\gamma\right)\right]\right.\\
\left.+\int_{\Lambda}^{\infty}\frac{kdk}{\left(2\pi\right)^{2}}\frac{\hbar^{2}k^{2}}{2M}\left[\frac{\sum_{\sigma}c_{a}^{(\sigma)}}{k^{2}}+\frac{\sum_{\sigma}\left(c_{b}^{(\sigma)}+Q_{cm}^{(\sigma)}\right)}{k^{4}}\right]\right\} ,\label{eq:3.22}
\end{multline}
where we have defined the corresponding two-body quantities $c_{a}^{(\sigma)}\equiv8\pi^{2}\mathcal{I}_{a}^{(\sigma)}$,
$c_{b}^{(\sigma)}\equiv-32\pi^{2}\mathcal{I}_{b}^{(\sigma)}$, and
$Q_{cm}^{(\sigma)}\equiv2\pi^{2}\int d{\bf R}\left(\nabla_{{\bf R}}\mathcal{A}_{\sigma}^{*}\cdot\nabla_{{\bf R}}\mathcal{A}_{\sigma}\right)$.
We can see that the divergent integral of $I_{\bar{\epsilon}}^{(2)}$
exactly compensates that of $I_{whole}^{(2)}$, and then the RHS of
Eq.(\ref{eq:3.2}), i.e., $I_{whole}^{(2)}-I_{\bar{s}_{\epsilon}}^{(2)}$
converges.\end{widetext}

The above procedure can easily be generalized to the many-body system
of $N$ spin-polarized fermions. The divergence of the corresponding
integral $I_{whole}^{(N)}$ arises when any two of fermions get close.
Since there are totally $\mathcal{N}=N\left(N-1\right)/2$ ways to
pair atoms, we obtain $I_{\bar{\mathcal{S}}_{\epsilon}}^{(N)}=\mathcal{N}I_{\bar{s}_{\epsilon}}^{(2)}$,
where the domain $\bar{\mathcal{S}}_{\epsilon}$ is the set of all
configurations $\left({\bf r}_{i},{\bf r}_{j}\right)$, in which $\left|{\bf r}_{i}-{\bf r}_{j}\right|<\epsilon$.
Finally, after redefining the constant $\mathcal{N}$ into the contacts,
the energy theorem for a many-body system can be rearranged as\begin{widetext}
\begin{multline}
E\left[1+\sum_{\sigma}\frac{\mathcal{C}_{a}^{(\sigma)}}{8\pi^{2}}\left(\ln\frac{2b_{\sigma}}{\epsilon}-\gamma\right)\right]=\lim_{\Lambda\rightarrow\infty}\left\{ \sum_{\left|{\bf k}\right|<\Lambda}\frac{\hbar^{2}k^{2}}{2M}n\left({\bf k}\right)\right.\\
\left.-\frac{\hbar^{2}}{16\pi^{2}M}\sum_{\sigma}\mathcal{C}_{a}^{(\sigma)}\left(\Lambda^{2}-\frac{\pi}{a_{\sigma}}\right)-\frac{\hbar^{2}}{8\pi^{2}M}\sum_{\sigma}\left[\frac{\mathcal{C}_{b}^{(\sigma)}}{2}\ln\left(\Lambda b_{\sigma}\right)+\left(\frac{\mathcal{C}_{b}^{(\sigma)}}{2}+\mathcal{Q}_{cm}^{(\sigma)}\right)\left(\ln\frac{\epsilon\Lambda}{2}+\gamma\right)\right]\right\} ,\label{eq:3.23}
\end{multline}
where $\gamma$ is the Euler's constant. Here, we should note that
unlike the $s$-wave case, the range $\epsilon$ of the $p$-wave
interaction appears in the energy theorem. This feature is resulted
from the non-normalizability of the higher-partial wave functions,
and the short-range physics becomes important, which has already been
pointed out in \cite{He2016C} for the three-dimensional systems.
The parameter $\Lambda$ in the expression of the limit characterizes
the cut-off of the momentum $k$ , and it screens off the details
of interatomic interactions from the outside. Therefore, the order
of $\Lambda$ should physically be larger than the inverse of the
interaction range $\epsilon$.\end{widetext}

\section{The high-frequency tail of the rf spectroscopy\label{sec:rf response}}

In the rf experiments, the fermions can be driven from the initially
occupied spin state $\left|g\right\rangle $ to an \emph{empty} spin
state $\left|e\right\rangle $, when the external rf field is tuned
near the transition frequency between the states $\left|g\right\rangle $
and $\left|e\right\rangle $. The universal scaling behavior at high
frequency of the rf response of the system is governed by contacts
\cite{Punk2007T,Schneider2010U,Braaten2010S,Yu2015U,Luciuk2016E}.
In this section, we are going to show how the contacts defined by
the adiabatic relations characterize such high-frequency scalings
of the rf spectroscopy of a spin-polarized Fermi gas in two dimensions.
Let us again start from a two-body picture, and consider two fermions
in the same spin state, which may simplify the presentation as much
as possible. The rf field is described in the momentum space by
\begin{equation}
\mathcal{H}_{rf}=\gamma_{rf}\sum_{{\bf k}}\left(e^{-i\omega t}c_{e{\bf k}}^{\dagger}c_{{\bf k}}+e^{i\omega t}c_{{\bf k}}^{\dagger}c_{e{\bf k}}\right),\label{eq:4.1}
\end{equation}
 where $\gamma_{rf}$ is the strength of the rf drive, $\omega$ is
the rf frequency, and $c_{e{\bf k}}^{\dagger}$ and $c_{{\bf k}}^{\dagger}$
are respectively the creation operators for fermions with the momentum
${\bf k}$ in the spin states $\left|e\right\rangle $ and $\left|g\right\rangle $.
The initial two-body state before the rf transition can be written
as
\begin{equation}
\left|\Psi_{i}\right\rangle =\frac{1}{\sqrt{2}}\sum_{{\bf k}_{1}{\bf k}_{2}}\tilde{\Psi}_{2D}\left({\bf k}_{1},{\bf k}_{2}\right)c_{{\bf k}_{1}}^{\dagger}c_{{\bf k}_{2}}^{\dagger}\left|0\right\rangle ,\label{eq:4.2}
\end{equation}
 where $\tilde{\Psi}_{2D}\left({\bf k}_{1},{\bf k}_{2}\right)$ is
the Fourier transform of the two-body wave function (\ref{eq:1.1})
(the variable ${\bf X}$ drops out in the two-body picture), i.e.,
\begin{equation}
\tilde{\Psi}_{2D}\left({\bf k}_{1},{\bf k}_{2}\right)=\int d{\bf r}_{1}d{\bf r}_{2}\Psi_{2D}\left({\bf r}_{1},{\bf r}_{2}\right)e^{-i{\bf k}_{1}\cdot{\bf r}_{1}}e^{-i{\bf k}_{2}\cdot{\bf r}_{2}}.\label{eq:4.3}
\end{equation}
 Acting Eq.(\ref{eq:4.1}) onto Eq.(\ref{eq:4.2}), we easily obtain
the final state after the rf transition,
\begin{equation}
\left|\Psi_{f}\right\rangle =\frac{\gamma_{rf}e^{-i\omega t}}{\sqrt{2}}\sum_{{\bf k}_{1}{\bf k}_{2}}\tilde{\Psi}_{2D}\left({\bf k}_{1},{\bf k}_{2}\right)\left(c_{e{\bf k}_{1}}^{\dagger}c_{{\bf k}_{2}}^{\dagger}-c_{e{\bf k}_{2}}^{\dagger}c_{{\bf k}_{1}}^{\dagger}\right)\left|0\right\rangle .\label{eq:4.4}
\end{equation}
 The physical meaning of Eq.(\ref{eq:4.4}) is quite apparent: after
the rf transition, there are two possible final states that one of
the two fermions is driven from the initial spin state $\left|g\right\rangle $
to the final spin state $\left|e\right\rangle $ with either momentum
${\bf k}_{1}$ or ${\bf k}_{2},$ and the probabilities are both $\gamma_{rf}^{2}\left|\tilde{\Psi}_{2D}\left({\bf k}_{1},{\bf k}_{2}\right)\right|^{2}/2$.
According to the Fermi's golden rule \cite{Yu2015U}, and taking these
two final states into account, the two-body rf transition rate takes
the form
\begin{equation}
\Gamma_{2}\left(\omega\right)=\frac{\pi\gamma_{rf}^{2}}{\hbar}\sum_{{\bf k}_{1}{\bf k}_{2}}\left|\tilde{\Psi}_{2D}\left({\bf k}_{1},{\bf k}_{2}\right)\right|^{2}\delta\left(\hbar\omega-\Delta E\right),\label{eq:4.5}
\end{equation}
 where $\Delta E$ is the energy difference between the final and
initial states. If the final spin state $\left|e\right\rangle $ has
an ignorable interaction with the initial spin state $\left|g\right\rangle $,
the final-state energy becomes
\begin{equation}
E_{f}=\frac{\hbar^{2}K^{2}}{4M}+\frac{\hbar^{2}k^{2}}{M}+\hbar\omega_{e}+\hbar\omega_{g},\label{eq:4.6}
\end{equation}
where ${\bf K}={\bf k}_{1}+{\bf k}_{2}$, ${\bf k}=\left({\bf k}_{1}-{\bf k}_{2}\right)/2$,
and $\omega_{e}$ and $\omega_{g}$ are the bare hyperfine frequencies
of the final and initial spin states, respectively. The energy of
the initial state with two fermions in the spin state $\left|g\right\rangle $
is
\begin{equation}
E_{i}=\frac{\hbar^{2}K^{2}}{4M}+\frac{\hbar^{2}q^{2}}{M}+2\hbar\omega_{g},\label{eq:4.7}
\end{equation}
 and $\hbar^{2}q^{2}/M$ is the relative energy of two fermions in
the spin state $\left|g\right\rangle $. Therefore, we have
\begin{equation}
\Delta E\approx\frac{\hbar^{2}k^{2}}{M}+\hbar\left(\omega_{e}-\omega_{g}\right)-\frac{\hbar^{2}q^{2}}{M},\label{eq:4.8}
\end{equation}
 then we obtain
\begin{equation}
\Gamma_{2}\left(\omega\right)=\frac{\pi\gamma_{rf}^{2}}{\hbar}\sum_{{\bf k}_{1}{\bf k}_{2}}\left|\tilde{\Psi}_{2D}\left({\bf k}_{1},{\bf k}_{2}\right)\right|^{2}\delta\left(\hbar\omega+\frac{\hbar^{2}\left(q^{2}-k^{2}\right)}{M}\right),\label{eq:4.9}
\end{equation}
 where we set the bare hyperfine splitting $\omega_{e}-\omega_{g}=0$
without loss of generality. Furthermore, if inserting Eq.(\ref{eq:4.3})
into Eq.(\ref{eq:4.9}), we may rewrite the rf transition rate as
\begin{equation}
\Gamma_{2}\left(\omega\right)=\frac{\pi\gamma_{rf}^{2}}{\hbar}\sum_{{\bf k}}n^{\prime}\left({\bf k}\right)\delta\left(\hbar\omega+\frac{\hbar^{2}\left(q^{2}-k^{2}\right)}{M}\right),\label{eq:4.10}
\end{equation}
 where
\begin{equation}
n^{\prime}\left({\bf k}\right)\equiv\int d{\bf R}\left|\int d{\bf r}\Psi_{2D}e^{-i{\bf k}\cdot{\bf r}}\right|^{2},\label{eq:4.11}
\end{equation}
and recall ${\bf R}=\left({\bf r}_{1}+{\bf r}_{2}\right)/2$ and ${\bf r}={\bf r}_{1}-{\bf r}_{2}$.
Comparing Eq.(\ref{eq:4.11}) with the definition of the single-particle
momentum distribution, i.e., Eq.(\ref{eq:2.1}), or specifically,
for a two-body system
\begin{equation}
n_{1}\left({\bf k}\right)=\int d{\bf r}_{2}\left|\int d{\bf r}_{1}\Psi_{2D}e^{-i{\bf k}\cdot{\bf r}_{1}}\right|^{2},\label{eq:4.12}
\end{equation}
we find $n^{\prime}\left({\bf k}\right)$ is not exactly the single-particle
momentum distribution of the system: $n^{\prime}\left({\bf k}\right)$
should have the same leading-order behavior as that of $n_{1}\left({\bf k}\right)$
at large ${\bf k}$, but different subleading-order behavior, in which
the c.m. contribution is excluded. As we can see from Eq.(\ref{eq:4.10}),
the high-frequency behavior of the rf transition rate is determined
by $n^{\prime}\left({\bf k}\right)$ at large ${\bf k}$ but still
smaller than $\epsilon^{-1}$, due to the delta function. Here, we
should carefully deal with the relative energy $\hbar^{2}q^{2}/M$
in the low-energy limit, since the subleading-order behavior becomes
important. Simply using the Fourier transform of the relative wave
function of two fermions at small ${\bf r}$, we easily obtain the
form of $n^{\prime}\left({\bf k}\right)$ at large ${\bf k}$, and
then the two-body rf transition rate $\Gamma_{2}\left(\omega\right)$
becomes
\begin{equation}
\Gamma_{2}\left(\omega\right)\approx\frac{M\gamma_{rf}^{2}}{16\pi\hbar^{3}}\left[\frac{\sum_{\sigma}c_{a}^{(\sigma)}}{M\omega/\hbar}+\frac{\sum_{\sigma}c_{b}^{(\sigma)}}{2\left(M\omega/\hbar\right)^{2}}\right]\label{eq:4.13}
\end{equation}
 at large $\omega$ but smaller than $\hbar/M\epsilon^{2}$, and again
$c_{a}^{(\sigma)}$, $c_{b}^{(\sigma)}$ are the two-body contacts.

For the many-body systems, all possible $\mathcal{N}$ pairs may contribute
to the high-frequency tail of the rf spectroscopy, when the two fermions
in them get close, while all the other fermions are far away. Therefore,
we can follow the above two-body route, and easily obtain the asymptotic
behavior of the rf response of the many-body system at large $\omega$,
after redefining the constant $\mathcal{N}$ into the contacts, i.e.,
\begin{equation}
\Gamma\left(\omega\right)\approx\frac{M\gamma_{rf}^{2}}{16\pi\hbar^{3}}\left[\frac{\sum_{\sigma}\mathcal{C}_{a}^{(\sigma)}}{M\omega/\hbar}+\frac{\sum_{\sigma}\mathcal{C}_{b}^{(\sigma)}}{2\left(M\omega/\hbar\right)^{2}}\right],\label{eq:4.14}
\end{equation}
 where $\mathcal{C}_{a}^{(\sigma)}$ and $\mathcal{C}_{b}^{(\sigma)}$
are corresponding many-body contacts, and $\Gamma\left(\omega\right)$
should obey the sum rule $\int d\omega\Gamma\left(\omega\right)=\pi\gamma_{rf}^{2}N/\hbar^{2}$
\cite{Braaten2010S}.

\section{Pair correlation function at short distances \label{sec:Pair-correlation-function}}

The pair correlation function $g_{2}\left({\bf s},{\bf t}\right)$
gives the probability of finding two fermions at positions ${\bf s}$
and ${\bf t}$ simultaneously, i.e., $g_{2}\left({\bf s},{\bf t}\right)\equiv\left\langle \hat{\rho}\left({\bf s}\right)\hat{\rho}\left({\bf t}\right)\right\rangle $,
where $\hat{\rho}\left({\bf s}\right)=\sum_{i}\delta\left({\bf s}-{\bf r}_{i}\right)$
is the density operator at the position ${\bf s}$. For a pure many-body
state $\left|\Psi_{2D}\right\rangle $ of $N$ fermions, we have
\begin{eqnarray}
g_{2}\left({\bf s},{\bf t}\right) & = & \int d{\bf r}_{1}d{\bf r}_{2}\cdots d{\bf r}_{N}\left\langle \Psi_{2D}\right|\hat{\rho}\left({\bf s}\right)\hat{\rho}\left({\bf t}\right)\left|\Psi_{2D}\right\rangle \nonumber \\
 & = & N\left(N-1\right)\int d{\bf X}\left|\Psi_{2D}\left({\bf X},{\bf R},{\bf r}\right)\right|^{2},\label{eq:5.1}
\end{eqnarray}
 where ${\bf R}=\left({\bf s}+{\bf t}\right)/2$, ${\bf r}={\bf s}-{\bf t}$,
and ${\bf X}$ denotes all the degrees of freedom of the fermions
except the ones at ${\bf s}$ and ${\bf t}$. Further more, we may
also integrate over the c.m. coordinate ${\bf R}$, and define the
spatially integrated pair correlation function as
\begin{equation}
G_{2}\left({\bf r}\right)\equiv\int d{\bf R}g_{2}\left({\bf R}+\frac{{\bf r}}{2},{\bf R}-\frac{{\bf r}}{2}\right).\label{5.2}
\end{equation}
 Using the asymptotic form the many-body wave function at short distance,
i.e., Eq.(\ref{eq:1.1}), we easily obtain
\begin{multline}
G_{2}\left({\bf r}\right)\approx N\left(N-1\right)\sum_{\sigma\sigma^{\prime}}\int d{\bf X}d{\bf R}\mathcal{A}_{\sigma^{\prime}}^{*}\mathcal{A}_{\sigma}\\
\left[\frac{1}{r^{2}}-\frac{q^{2}}{2}\left(\ln\frac{r}{2b_{\sigma^{\prime}}}+\ln\frac{r}{2b_{\sigma}}\right)\right]\Omega_{\sigma^{\prime}}^{*}\left(\varphi\right)\Omega_{\sigma}\left(\varphi\right).\label{eq:5.3}
\end{multline}
 If we are only interested in the dependence of $G_{2}\left({\bf r}\right)$
on $r=\left|{\bf r}\right|$, we may integrate $G_{2}\left({\bf r}\right)$
over the direction of ${\bf r}$, and obtain
\begin{eqnarray}
G_{2}\left(r\right) & \approx & N\left(N-1\right)\sum_{\sigma}\left(\frac{\mathcal{I}_{a}^{(\sigma)}}{r^{2}}+2\mathcal{I}_{b}^{(\sigma)}\ln\frac{r}{2b_{\sigma}}\right)\nonumber \\
 & = & \frac{1}{4\pi^{2}}\sum_{\sigma}\left(\frac{\mathcal{C}_{a}^{(\sigma)}}{r^{2}}-\frac{\mathcal{C}_{b}^{(\sigma)}}{2}\ln\frac{r}{2b_{\sigma}}\right).\label{eq:5.4}
\end{eqnarray}
 We can see that the short-distance behavior of the pair correlation
function of a spin-polarized Fermi gas is also completely captured
by the $p$-wave contacts $\mathcal{C}_{a}^{(\sigma)}$ and $\mathcal{C}_{b}^{(\sigma)}$.

\section{Generalized virial theorem and pressure relation\label{sec:Generalized-virial-theorem}}

Let us consider a spin-polarized Fermi gas trapped in the harmonic
potential $V$, then the Helmholtz free energy $F$ should be the
function of the temperature $T$, the trap frequency $\omega$, the
atom number $N$, and the interatomic $p$-wave interaction strength
characterized by the 2D scattering area $a_{\sigma}$ as well as the
effective range $b_{\sigma}$, i.e., $F\left(T,\omega,a_{\sigma},b_{\sigma},N\right)$.
The generalized virial theorem can be obtained according to the dimensional
analysis \cite{Braaten2008E,Zhang2009U,Barth2011T}. Using $\hbar\omega$
as the unit of the energy, the Helmholtz free energy may be written
as
\begin{equation}
F\left(T,\omega,a_{\sigma},b_{\sigma},N\right)=\hbar\omega f\left(\frac{k_{B}T}{\hbar\omega},\frac{\hbar^{2}/Ma_{\sigma}}{\hbar\omega},\frac{\hbar^{2}/Mb_{\sigma}^{2}}{\hbar\omega},N\right),\label{eq:6.1}
\end{equation}
 where the function $f$ is just a dimensionless function, and $k_{B}$
is the Boltzmann constant. Then the free energy $F$ should have the
following scaling property,
\begin{equation}
F\left(\lambda T,\lambda\omega,\lambda^{-1}a_{\sigma},\lambda^{-1/2}b_{\sigma},N\right)=\lambda F\left(T,\omega,a_{\sigma},b_{\sigma},N\right).\label{eq:6.2}
\end{equation}
Taking the derivative with respect to $\lambda$ on both sides of
Eq.(\ref{eq:6.2}), and then setting $\lambda=1$, we obtain
\begin{equation}
\left(T\frac{\partial}{\partial T}+\omega\frac{\partial}{\partial\omega}-a_{\sigma}\frac{\partial}{\partial a_{\sigma}}-\frac{b_{\sigma}}{2}\frac{\partial}{\partial b_{\sigma}}\right)F=F.\label{eq:6.3}
\end{equation}
Since the Helmholtz free energy is just the Legendre transform of
the energy, its partial derivatives at constant $T$ with respect
to $\omega$, $a_{\sigma}$, and $b_{\sigma}$ are equal to those
of the energy at the associated value of the entropy $S$. Combining
the adiabatic relations (\ref{eq:2.18}) and (\ref{eq:2.19}), and
$dF=dE-SdT$, we easily obtain
\begin{equation}
E=2\left\langle V\right\rangle -\frac{\hbar^{2}\mathcal{C}_{a}^{(\sigma)}}{16\pi Ma_{\sigma}}-\frac{\hbar^{2}\mathcal{C}_{b}^{(\sigma)}}{32\pi^{2}M}.\label{eq:6.4}
\end{equation}

The pressure relation can be derived following the similar route.
Let us consider the free energy density $\mathcal{F}$, which has
the dimension of (energy)$^{2}$ up to the factors $\hbar$ and $M$.
Assuming $\kappa$ is an arbitrary quantity with dimension of (energy)$^{1}$,
the free energy density can be written as
\begin{equation}
\mathcal{F}\left(T,a_{\sigma},b_{\sigma},n\right)=\frac{M\kappa^{2}}{\hbar^{2}}f\left(\frac{k_{B}T}{\kappa},\frac{\hbar^{2}/Ma_{\sigma}}{\kappa},\frac{\hbar^{2}/Mb_{\sigma}^{2}}{\kappa},\frac{\hbar^{2}n/M}{\kappa}\right),\label{eq:6.5}
\end{equation}
 where $n$ is the atom density. Then we have
\begin{equation}
\mathcal{F}\left(\lambda T,\lambda^{-1}a_{\sigma},\lambda^{-1/2}b_{\sigma},\lambda n\right)=\lambda^{2}\mathcal{F}\left(T,a_{\sigma},b_{\sigma},n\right),\label{eq:6.6}
\end{equation}
 which similarly yields
\begin{equation}
\left(T\frac{\partial}{\partial T}-a_{\sigma}\frac{\partial}{\partial a_{\sigma}}-\frac{b_{\sigma}}{2}\frac{\partial}{\partial b_{\sigma}}+n\frac{\partial}{\partial n}\right)\mathcal{F}=2\mathcal{F}.\label{eq:6.7}
\end{equation}
 Combining $P=-\mathcal{F}+n\mu$, where $\mu$ is the chemical potential,
and the adiabatic relations, we finally obtain the pressure relation
\begin{equation}
P=\varepsilon+\frac{\hbar^{2}\mathcal{C}_{a}^{(\sigma)}}{16\pi Ma_{\sigma}}+\frac{\hbar^{2}\mathcal{C}_{b}^{(\sigma)}}{32\pi^{2}M},\label{eq:6.8}
\end{equation}
 where $\varepsilon$ is the energy density of the system.

\section{Conclusions\label{sec:Conclusions}}

To conclude, we have systematically studied the full set of universal
relations of a two-dimensional spin-polarized Fermi gas with $p$-wave
interactions. If the $p$-wave contacts are defined according to the
adiabatic relations, we find that the universal relations of the system,
such as the high-frequency tail of the radio-frequency response, short-distance
behavior of the pair correlation function, generalized virial theorem,
and pressure relation are fully captured by the contacts we define.
As we anticipate, an extra term resulted from the center-of-mass motions
of the pairs appears in the subleading tail ($k^{-4}$) of the large
momentum distribution besides the contact related to the effective
range, similar to what happens in a three-dimensional $p$-wave Fermi
gas. Furthermore, such an extra term results in an additional divergence
for the energy theorem, which should carefully be handled with. We
show that all the divergences of the kinetic energy are exactly compensated
by the interatomic interaction energy, and the total internal energy
of the system converges. Our results could easily be generalized for
higher-partial-wave scatterings. The predicted universal relations
could readily be confirmed in current cold-atom experiments with spin-polarized
Fermi gases of $^{40}$K and $^{6}$Li atoms.
\begin{acknowledgments}
We gratefully acknowledge valuable discussions with Hui Hu, Xia-Ji
Liu, Zhenhua Yu, Qi Zhou and Shina Tan. This work has been supported
by the NKRDP (National Key Research and Development Program) under
Grant No.2016YFA0301503, NSFC (Grant No.11474315,11674358, 11434015),
CAS under Grant No.YJKYYQ20170025, and the Strategic Priority Research
Program of the Chinese Academy of Sciences under Grant No. XDB21010100.
\end{acknowledgments}

\begin{widetext}

\appendix

\section{the normalization of the wave function\label{sec:TheNormalization}}

In this appendix, we are going to discuss the normalization of the
many-body wave function $\Psi_{2D}$ in two dimension, and calculate
$\int_{\mathcal{S}_{\epsilon}}\prod_{i=1}^{N}d{\bf r}_{i}\left|\Psi_{2D}\right|^{2}$.
The similar normalization has been discussed for a three-dimensional
system (see Appendix A of \cite{Peng2016L}), and such normalization
is related to the probability of finding two fermions inside the interaction
range $\epsilon$. Let us consider $N$ spin-polarized fermions, and
when any two of them interact with each other, for example, fermions
$i$ and $j$, all the others are far away. For two different relative
energies of the pair $\left(i,j\right)$, i.e., $\mathcal{E}$ and
$\mathcal{E}^{\prime}$, due to the orthogonality of the wave function,
we have $\int d{\bf X}d{\bf R}d{\bf r}\Psi_{2D}^{\prime*}\Psi_{2D}=0$.
Then from the Schr\"{o}dinger equations satisfied by $\Psi_{2D}^{\prime}$
and $\Psi_{2D}$, we find the probability of finding the pair $\left(i,j\right)$
inside the interaction range $\epsilon$ should be
\begin{eqnarray}
\int_{r<\epsilon}d{\bf X}d{\bf R}d{\bf r}\left|\Psi_{2D}\right|^{2} & = & -\lim_{\mathcal{E}^{\prime}\rightarrow\mathcal{E}}\int_{r>\epsilon}d{\bf X}d{\bf R}d{\bf r}\Psi_{2D}^{\prime*}\Psi_{2D}\nonumber \\
 & = & -\lim_{\mathcal{E}^{\prime}\rightarrow\mathcal{E}}\frac{\hbar^{2}/M}{\mathcal{E}-\mathcal{E}^{\prime}}\int_{r=\epsilon}d{\bf X}d{\bf R}d{\bf r}\left(\Psi_{2D}^{\prime*}\frac{\partial}{\partial r}\Psi_{2D}-\Psi_{2D}\frac{\partial}{\partial r}\Psi_{2D}^{\prime*}\right)\nonumber \\
 & = & -\lim_{\mathcal{E}^{\prime}\rightarrow\mathcal{E}}\sum_{\sigma\sigma^{\prime}}\int d{\bf X}d{\bf R}\mathcal{A}_{\sigma^{\prime}}^{\prime*}\mathcal{A}_{\sigma}\cdot\frac{\hbar^{2}/M}{\mathcal{E}-\mathcal{E}^{\prime}}\int_{0}^{2\pi}\epsilon d\varphi\left(\psi_{\sigma^{\prime}}^{\prime*}\frac{\partial}{\partial r}\psi_{\sigma}-\psi_{\sigma}\frac{\partial}{\partial r}\psi_{\sigma^{\prime}}^{\prime*}\right)\nonumber \\
 & = & -\sum_{\sigma}\mathcal{I}_{a}^{(\sigma)}\left[-\left(\gamma+\ln\frac{\epsilon}{2}\right)+\frac{\partial}{\partial q^{2}}\left(\frac{\pi}{2}q^{2}\cot\delta_{\sigma}-q^{2}\ln q\right)\right]\nonumber \\
 & = & -\sum_{\sigma}\mathcal{I}_{a}^{(\sigma)}\left(\ln\frac{2b_{\sigma}}{\epsilon}-\gamma\right),\label{eq:A1}
\end{eqnarray}
where we have used the effective-range expansion of the scattering
phase shift (\ref{eq:1.7}), $q^{2}=M\mathcal{E}/\hbar^{2}$, and
$\gamma$ is the Euler's constant. We can see that the bound for the
effective range $b_{\sigma}$ exists, i.e., $b_{\sigma}<\epsilon e^{\gamma}/2$,
in order to guarantee the positive probability of finding two atoms
inside the interaction range. This is an alternative expression of
the Wigner's bound on the effective range for the $p$-wave interaction
in two dimensions \cite{Peng2016L,Wigner1955L,Phillips1997H,Hammer2009C}.
Then the total probability of finding any pair of fermions inside
the interaction range is
\begin{equation}
\int_{\bar{\mathcal{S}}_{\epsilon}}\prod_{i=1}^{N}d{\bf r}_{i}\left|\Psi_{2D}\right|^{2}=\frac{N\left(N-1\right)}{2}\int_{r<\epsilon}d{\bf X}d{\bf R}d{\bf r}\left|\Psi_{2D}\right|^{2}=-\mathcal{N}\sum_{\sigma}\mathcal{I}_{a}^{(\sigma)}\left(\ln\frac{2b_{\sigma}}{\epsilon}-\gamma\right),\label{eq:A2}
\end{equation}
 where $\mathcal{I}_{a}^{(\sigma)}$ is defined in Eq.(\ref{eq:1.10}),
and $\bar{\mathcal{S}}_{\epsilon}$ is the set of all configurations
that there is only one pair inside the interaction range. Consequently,
we obtain
\begin{equation}
\int_{\mathcal{S}_{\epsilon}}\prod_{i=1}^{N}d{\bf r}_{i}\left|\Psi_{2D}\right|^{2}=1+\mathcal{N}\sum_{\sigma}\mathcal{I}_{a}^{(\sigma)}\left(\ln\frac{2b_{\sigma}}{\epsilon}-\gamma\right).\label{eq:A3}
\end{equation}

\section{Derivation details of the momentum distribution\label{sec:MomentumDistribution}}

In this part of the appendix, we present the calculation details in
the derivation from Eq.(\ref{eq:2.14}) to (\ref{eq:2.15}). The integral
over the direction of ${\bf k}$ for the leading-order term, i.e.,
$\sim k^{-2}$, can easily be obtained, and the coefficient takes
the simple form of $\sum_{\sigma}8\pi^{2}\mathcal{N}\mathcal{I}_{a}^{(\sigma)}$,
which then we define as $\sum_{\sigma}\mathcal{C}_{a}^{(\sigma)}$.
For the $k^{-3}$-order term, we find
\begin{eqnarray}
\tau_{3} & \equiv & \text{{\bf Im}}\mathcal{N}\int d{\bf X}d{\bf R}\sum_{\sigma\sigma^{\prime}}\mathcal{A}_{\sigma}e^{i\sigma\varphi_{{\bf k}}}\alpha_{\sigma^{\prime}}^{*}\cdot\frac{4\pi}{k^{3}}\nonumber \\
 & = & -4\pi\mathcal{N}\text{{\bf Im}}\int d{\bf X}d{\bf R}\sum_{\sigma\sigma^{\prime}}\mathcal{A}_{\sigma}\left[\left(\nabla_{{\bf R}}\mathcal{A}_{\sigma^{\prime}}^{*}\cdot\hat{{\bf k}}\right)+i\sigma^{\prime}\left(\nabla_{{\bf R}}\mathcal{A}_{\sigma^{\prime}}^{*}\cdot\hat{{\bf \varphi}}_{{\bf k}}\right)\right]\cdot\frac{e^{i\left(\sigma-\sigma^{\prime}\right)\varphi_{{\bf k}}}}{k^{3}},\label{eq:B1}
\end{eqnarray}
 where $\hat{{\bf k}}$, $\hat{\varphi}_{{\bf k}}$ are respectively
the unit vectors of the radial and azimuthal directions of ${\bf k}$.
Obviously, $\tau_{3}$ is simply the linear combination of $e^{i\left(\sigma-\sigma^{\prime}\right)\varphi_{{\bf k}}}\sin\varphi_{{\bf k}}$
and $e^{i\left(\sigma-\sigma^{\prime}\right)\varphi_{{\bf k}}}\cos\varphi_{{\bf k}}$,
which automatically vanishes if integrating over $\varphi_{{\bf k}}$.

Let us look at the $k^{-4}$-order term, which includes two terms.
The first term becomes $-32\pi^{2}\sum_{\sigma}\mathcal{N}\mathcal{I}_{b}^{(\sigma)}/k^{4}$
if integrating over $\varphi_{{\bf k}}$, and then we define as $\sum_{\sigma}\mathcal{C}_{b}^{(\sigma)}/k^{4}$.
As to the second term, we rewrite it as
\begin{eqnarray}
\chi & = & \pi\mathcal{N}\int d{\bf X}d{\bf R}\sum_{\sigma\sigma^{\prime}}\left[\nabla_{{\bf R}}\mathcal{A}_{\sigma^{\prime}}^{*}\cdot\nabla_{{\bf k}}f_{\sigma^{\prime}}^{*}\right]\left[\nabla_{{\bf R}}\mathcal{A}_{\sigma}\cdot\nabla_{{\bf k}}f_{\sigma}\right]\nonumber \\
 &  & -\frac{\pi}{2}\mathcal{N}\int d{\bf X}d{\bf R}\sum_{\sigma\sigma^{\prime}}\left\{ \mathcal{A}_{\sigma^{\prime}}^{*}f_{\sigma^{\prime}}^{*}\left(\nabla_{{\bf R}}\cdot\nabla_{{\bf k}}\right)\left[\nabla_{{\bf R}}\mathcal{A}_{\sigma}\cdot\nabla_{{\bf k}}f_{\sigma}\right]+\mathcal{A}_{\sigma}f_{\sigma}\left(\nabla_{{\bf R}}\cdot\nabla_{{\bf k}}\right)\left[\nabla_{{\bf R}}\mathcal{A}_{\sigma^{\prime}}^{*}\cdot\nabla_{{\bf k}}f_{\sigma^{\prime}}^{*}\right]\right\} ,\label{eq:B2}
\end{eqnarray}
 where we have defined $f_{\sigma}\left({\bf k}\right)\equiv e^{i\sigma\varphi_{{\bf k}}}/k$.
Since the function $\mathcal{A}_{\sigma}$ is regular and should either
decay to zero at infinity or satisfy a periodic boundary condition
in a box \cite{Tan2008}, after partially integrating, $\chi$ becomes
\begin{eqnarray}
\chi & = & \pi\mathcal{N}\int d{\bf X}d{\bf R}\sum_{\sigma\sigma^{\prime}}\left[\nabla_{{\bf R}}\mathcal{A}_{\sigma^{\prime}}^{*}\cdot\nabla_{{\bf k}}f_{\sigma^{\prime}}^{*}\right]\left[\nabla_{{\bf R}}\mathcal{A}_{\sigma}\cdot\nabla_{{\bf k}}f_{\sigma}\right]\nonumber \\
 &  & +\frac{\pi}{2}\mathcal{N}\int d{\bf X}d{\bf R}\sum_{\sigma\sigma^{\prime}}\left\{ f_{\sigma^{\prime}}^{*}\left(\nabla_{{\bf R}}\mathcal{A}_{\sigma^{\prime}}^{*}\cdot\nabla_{{\bf k}}\right)\left[\nabla_{{\bf R}}\mathcal{A}_{\sigma}\cdot\nabla_{{\bf k}}f_{\sigma}\right]+f_{\sigma}\left(\nabla_{{\bf R}}\mathcal{A}_{\sigma}\cdot\nabla_{{\bf k}}\right)\left[\nabla_{{\bf R}}\mathcal{A}_{\sigma^{\prime}}^{*}\cdot\nabla_{{\bf k}}f_{\sigma^{\prime}}^{*}\right]\right\} \nonumber \\
 & = & \frac{\pi}{2}\mathcal{N}\int d{\bf X}d{\bf R}\sum_{\sigma\sigma^{\prime}}\sum_{ij}\frac{\partial\mathcal{A}_{\sigma^{\prime}}^{*}}{\partial R_{i}}\frac{\partial\mathcal{A}_{\sigma}}{\partial R_{j}}\left(2\frac{\partial f_{\sigma^{\prime}}^{*}}{\partial k_{i}}\frac{\partial f_{\sigma}}{\partial k_{j}}+f_{\sigma^{\prime}}^{*}\frac{\partial^{2}f_{\sigma}}{\partial k_{i}\partial k_{j}}+f_{\sigma}\frac{\partial^{2}f_{\sigma^{\prime}}^{*}}{\partial k_{j}\partial k_{i}}\right)\nonumber \\
 & = & \frac{\pi}{2}\mathcal{N}\int d{\bf X}d{\bf R}\sum_{\sigma\sigma^{\prime}}\sum_{ij}\frac{\partial\mathcal{A}_{\sigma^{\prime}}^{*}}{\partial R_{i}}\frac{\partial\mathcal{A}_{\sigma}}{\partial R_{j}}\left[\frac{\partial}{\partial k_{i}}\left(f_{\sigma^{\prime}}^{*}\frac{\partial f_{\sigma}}{\partial k_{j}}\right)+\frac{\partial}{\partial k_{j}}\left(f_{\sigma}\frac{\partial f_{\sigma^{\prime}}^{*}}{\partial k_{i}}\right)\right],\label{eq:B3}
\end{eqnarray}
where the indices $i,j$ denote $\left\{ x,y,z\right\} $. Inserting
\begin{equation}
f_{\sigma}\left({\bf k}\right)=\frac{e^{i\sigma\varphi_{{\bf k}}}}{k}=\frac{k_{x}+i\sigma k_{y}}{k^{2}},\label{eq:B4}
\end{equation}
 into Eq.(\ref{eq:B3}), and integrating over $\varphi_{{\bf k}}$
by using $k_{x}=k\cos\varphi_{{\bf k}}$ and $k_{y}=k\sin\varphi_{{\bf k}}$,
we arrive at
\begin{eqnarray}
\chi & = & \frac{2\pi^{2}}{k^{4}}\mathcal{N}\int d{\bf X}d{\bf R}\sum_{\sigma}\left(\frac{\partial\mathcal{A}_{\sigma}^{*}}{\partial R_{x}}\frac{\partial\mathcal{A}_{\sigma}}{\partial R_{x}}-i\sigma\frac{\partial\mathcal{A}_{\sigma}^{*}}{\partial R_{x}}\frac{\partial\mathcal{A}_{\sigma}}{\partial R_{y}}+i\sigma\frac{\partial\mathcal{A}_{\sigma}^{*}}{\partial R_{y}}\frac{\partial\mathcal{A}_{\sigma}}{\partial R_{x}}+\frac{\partial\mathcal{A}_{\sigma}^{*}}{\partial R_{y}}\frac{\partial\mathcal{A}_{\sigma}}{\partial R_{y}}\right)\nonumber \\
 & = & 2\pi^{2}\sum_{\sigma}\mathcal{N}\int d{\bf X}d{\bf R}\left(\nabla_{{\bf R}}\mathcal{A}_{\sigma}^{*}\cdot\nabla_{{\bf R}}\mathcal{A}_{\sigma}\right)\cdot\frac{1}{k^{4}},\label{eq:B5}
\end{eqnarray}
where we have used
\begin{equation}
\int d{\bf R}\left(\frac{\partial\mathcal{A}_{\sigma}^{*}}{\partial R_{x}}\frac{\partial\mathcal{A}_{\sigma}}{\partial R_{y}}-\frac{\partial\mathcal{A}_{\sigma}^{*}}{\partial R_{y}}\frac{\partial\mathcal{A}_{\sigma}}{\partial R_{x}}\right)=0.\label{eq:B6}
\end{equation}

\section{Calculation of $\nabla_{{\bf r}}^{2}\left[N_{m}\left(qr\right)\Omega_{m}^{(\sigma)}\left(\varphi\right)\right]$\label{sec:NormOfBesselFunction}}

The Bessel function of the second kind $N_{m}\left(x\right)$ takes
the following series power form at small $x$ \cite{Abramowitz1972H},
\begin{equation}
N_{m}\left(x\right)=-\frac{1}{\pi}\sum_{s=0}^{m-1}\frac{\left(m-s-1\right)!}{s!})\left(\frac{x}{2}\right)^{2s-m}+\frac{2}{\pi}\ln\left(\frac{x}{2}\right)J_{m}\left(x\right)-\frac{1}{\pi}\sum_{s=0}^{\infty}\left(-\right)^{s}\frac{\psi\left(s\right)+\psi\left(s+m+1\right)}{s!\left(s+m\right)!}\left(\frac{x}{2}\right)^{2s+m},\label{eq:C1}
\end{equation}
 where $J_{m}\left(x\right)$ is the Bessel function of the first
kind, and $\psi\left(\cdot\right)$ is the digamma function. Using
the form of $\nabla_{{\bf r}}^{2}$ in the polar coordinate, we find
\begin{equation}
\nabla_{{\bf r}}^{2}\left[N_{m}\left(qr\right)\Omega_{m}^{(\sigma)}\left(\varphi\right)\right]=\left[\left(\frac{1}{r}\frac{\partial}{\partial r}r\frac{\partial}{\partial r}-\frac{m^{2}}{r^{2}}\right)N_{m}\left(qr\right)\right]\Omega_{m}^{(\sigma)}\left(\varphi\right).\label{eq:C2}
\end{equation}
For $m\ge1$, we may separate the most singular term of $N_{m}\left(x\right)$
at small $x$, i.e., $-2^{m}\left(m-1\right)!/\pi x^{m}$, and find
\begin{equation}
\left(\frac{1}{x}\frac{\partial}{\partial x}x\frac{\partial}{\partial x}-\frac{m^{2}}{x^{2}}\right)\left[N_{m}\left(x\right)+\frac{2^{m}}{\pi}\cdot\frac{\left(m-1\right)!}{x^{m}}\right]=-N_{m}\left(x\right).\label{eq:C3}
\end{equation}
 Subsequently, we obtain
\begin{equation}
\nabla_{{\bf r}}^{2}\left[N_{m}\left(qr\right)\Omega_{m}^{(\sigma)}\left(\varphi\right)\right]=\left\{ \left(\frac{1}{r}\frac{\partial}{\partial r}r\frac{\partial}{\partial r}-\frac{m^{2}}{r^{2}}\right)\left[-\frac{2^{m}}{\pi}\cdot\frac{\left(m-1\right)!}{\left(qr\right)^{m}}\right]-q^{2}N_{m}\left(qr\right)\right\} \Omega_{m}^{(\sigma)}\left(\varphi\right).\label{eq:C4}
\end{equation}
Using \cite{Estrada1989R,Idziaszek2009A}
\begin{equation}
\frac{d}{dr}\frac{1}{r^{m}}=-\frac{m}{r^{m+1}}+\frac{\left(-\right)^{m}}{m!}\delta^{(m)}\left(r\right),\label{eq:C5}
\end{equation}
 where $\delta^{(m)}\left(r\right)$ is the $n$-th derivative of
the Dirac delta function $\delta\left(r\right)$, and $\delta^{(m)}\left(r\right)=\left(-\right)^{m}m!\delta\left(r\right)/r^{m}$,
we obtain
\begin{equation}
\left(\frac{1}{r}\frac{\partial}{\partial r}r\frac{\partial}{\partial r}-\frac{m^{2}}{r^{2}}\right)\left(\frac{1}{r^{m}}\right)=-\frac{2m}{r^{m+1}}\delta\left(r\right),\label{eq:C6}
\end{equation}
and then
\begin{equation}
\nabla_{{\bf r}}^{2}\left[N_{m}\left(qr\right)\Omega_{m}^{(\sigma)}\left(\varphi\right)\right]=\left[\frac{2^{m+1}}{\pi}\cdot\frac{m!}{q^{m}}\frac{\delta\left(r\right)}{r^{m+1}}-q^{2}N_{m}\left(qr\right)\right]\Omega_{m}^{(\sigma)}\left(\varphi\right),\label{eq:C7}
\end{equation}
 which finally yields Eq.(\ref{eq:3.16}).

\section{Calculation details of Eqs.(\ref{eq:3.19}) and (\ref{eq:3.21})\label{sec:Calculation-details-of}}

In this part of the appendix, we present the calculation details of
the derivations of Eqs.(\ref{eq:3.19}) and (\ref{eq:3.21}). Let
us look at Eq.(\ref{eq:3.19}) first. Using the identity (\ref{eq:3.16}),
we easily find
\begin{align}
 & \frac{\pi^{2}\hbar^{2}q^{2}}{4M}\int_{r<\epsilon}d{\bf r}N_{1}\left(qr\right)\Omega_{1}^{(\sigma^{\prime})*}\left(\varphi\right)\nabla_{{\bf r}}^{2}\left[N_{1}\left(qr\right)\Omega_{1}^{(\sigma)}\left(\varphi\right)\right]\nonumber \\
= & \frac{\pi\hbar^{2}q}{M}\int_{r<\epsilon}d{\bf r}\left[N_{1}\left(qr\right)\Omega_{1}^{(\sigma^{\prime})*}\left(\varphi\right)\right]\left[\frac{\delta\left(r\right)}{r^{2}}\Omega_{1}^{(\sigma)}\left(\varphi\right)\right]-\frac{\pi^{2}\hbar^{2}q^{4}}{4M}\int_{r<\epsilon}d{\bf r}\left[N_{1}\left(qr\right)\Omega_{1}^{(\sigma^{\prime})*}\left(\varphi\right)\right]\left[N_{1}\left(qr\right)\Omega_{1}^{(\sigma)}\left(\varphi\right)\right].\label{eq:D1}
\end{align}
Obviously, the two integrals of Eq.(\ref{eq:D1}) are both divergent.
Since we know
\begin{equation}
N_{1}\left(qr\right)=-\frac{2}{\pi qr}+\frac{qr}{\pi}\left(\ln\frac{qr}{2}+\gamma-\frac{1}{2}\right)+O\left(qr\right)^{3}\label{eq:D2}
\end{equation}
at $qr\sim0$, the first integral of Eq.(\ref{eq:D1}) becomes
\begin{multline}
\frac{\pi\hbar^{2}q}{M}\int_{r<\epsilon}d{\bf r}\left[N_{1}\left(qr\right)\Omega_{1}^{(\sigma^{\prime})*}\left(\varphi\right)\right]\left[\frac{\delta\left(r\right)}{r^{2}}\Omega_{1}^{(\sigma)}\left(\varphi\right)\right]\\
=-\frac{2\hbar^{2}}{M}\int_{r<\epsilon}d{\bf r}\frac{\Omega_{1}^{(\sigma^{\prime})*}\left(\varphi\right)}{r}\left[\frac{\delta\left(r\right)}{r^{2}}\Omega_{1}^{(\sigma)}\left(\varphi\right)\right]+\frac{\hbar^{2}q^{2}}{M}\int_{r<\epsilon}d{\bf r}\left[r\ln\frac{qr}{2}\Omega_{1}^{(\sigma^{\prime})*}\left(\varphi\right)\right]\left[\frac{\delta\left(r\right)}{r^{2}}\Omega_{1}^{(\sigma)}\left(\varphi\right)\right]+\delta_{\sigma\sigma^{\prime}}\frac{\hbar^{2}q^{2}}{2M}\left(2\gamma-1\right).\label{eq:D3}
\end{multline}
 Using the Fourier transform of $\frac{\delta\left(r\right)}{r^{2}}\Omega_{1}^{(\sigma)}\left(\varphi\right)$,
i.e.,
\begin{equation}
\mathcal{F}\left[\frac{\delta\left(r\right)}{r^{2}}\Omega_{1}^{(\sigma)}\left(\varphi\right)\right]=-i\sqrt{2\pi}e^{i\sigma\varphi_{{\bf k}}}\frac{k}{2},\label{eq:D4}
\end{equation}
 we obtain
\begin{eqnarray}
-\frac{2\hbar^{2}}{M}\int_{r<\epsilon}d{\bf r}\frac{\Omega_{1}^{(\sigma^{\prime})*}\left(\varphi\right)}{r}\left[\frac{\delta\left(r\right)}{r^{2}}\Omega_{1}^{(\sigma)}\left(\varphi\right)\right] & = & -\delta_{\sigma\sigma^{\prime}}\frac{4\pi^{2}\hbar^{2}}{M}\int\frac{kdk}{\left(2\pi\right)^{2}}\left[1-J_{0}\left(k\epsilon\right)\right]\nonumber \\
 & = & -\delta_{\sigma\sigma^{\prime}}\frac{4\pi^{2}\hbar^{2}}{M}\int\frac{kdk}{\left(2\pi\right)^{2}}\nonumber \\
 & = & -\delta_{\sigma\sigma^{\prime}}\frac{4\pi^{2}\hbar^{2}}{M}\lim_{\Lambda\rightarrow\infty}\left[\int_{0}^{\Lambda}\frac{kdk}{\left(2\pi\right)^{2}}+\int_{\Lambda}^{\infty}\frac{kdk}{\left(2\pi\right)^{2}}\right]\nonumber \\
 & = & \delta_{\sigma\sigma^{\prime}}\lim_{\Lambda\rightarrow\infty}\left[-\frac{\hbar^{2}\Lambda^{2}}{2M}-8\pi^{2}\int_{\Lambda}^{\infty}\frac{kdk}{\left(2\pi\right)^{2}}\frac{\hbar^{2}k^{2}}{2M}\frac{1}{k^{2}}\right].\label{eq:D5}
\end{eqnarray}
 Similarly,
\begin{multline}
\frac{\hbar^{2}q^{2}}{M}\int_{r<\epsilon}d{\bf r}\left[r\ln\frac{qr}{2}\Omega_{1}^{(\sigma^{\prime})*}\left(\varphi\right)\right]\left[\frac{\delta\left(r\right)}{r^{2}}\Omega_{1}^{(\sigma)}\left(\varphi\right)\right]=\delta_{\sigma\sigma^{\prime}}\frac{2\pi^{2}\hbar^{2}q^{2}}{M}\int\frac{kdk}{\left(2\pi\right)^{2}}\\
\frac{-2\left[1-J_{0}\left(k\epsilon\right)\right]+k\epsilon J_{1}\left(k\epsilon\right)+\left[2J_{1}\left(k\epsilon\right)-k\epsilon J_{0}\left(k\epsilon\right)\right]k\epsilon\ln\frac{\epsilon q}{2}}{k^{2}}.\label{eq:D6}
\end{multline}
 Since
\begin{align}
 & \frac{2\pi^{2}\hbar^{2}q^{2}}{M}\int\frac{kdk}{\left(2\pi\right)^{2}}\frac{-2\left[1-J_{0}\left(k\epsilon\right)\right]}{k^{2}}\nonumber \\
= & \frac{2\pi^{2}\hbar^{2}q^{2}}{M}\lim_{\Lambda\rightarrow\infty}\left[\int_{0}^{\Lambda}\frac{kdk}{\left(2\pi\right)^{2}}\frac{-2+2J_{0}\left(k\epsilon\right)}{k^{2}}+\int_{\Lambda}^{\infty}\frac{kdk}{\left(2\pi\right)^{2}}\frac{-2+2J_{0}\left(k\epsilon\right)}{k^{2}}\right]\nonumber \\
= & \lim_{\Lambda\rightarrow\infty}\left[-\frac{\hbar^{2}q^{2}}{M}\left(\gamma+\ln\frac{\epsilon\Lambda}{2}\right)-8\pi^{2}q^{2}\int_{\Lambda}^{\infty}\frac{kdk}{\left(2\pi\right)^{2}}\frac{\hbar^{2}k^{2}}{2M}\frac{1}{k^{4}}\right],\label{eq:D7}
\end{align}
and
\begin{equation}
\frac{2\pi^{2}\hbar^{2}q^{2}}{M}\int\frac{kdk}{\left(2\pi\right)^{2}}\frac{k\epsilon J_{1}\left(k\epsilon\right)+\left[2J_{1}\left(k\epsilon\right)-k\epsilon J_{0}\left(k\epsilon\right)\right]k\epsilon\ln\frac{q\epsilon}{2}}{k^{2}}=\frac{\hbar^{2}q^{2}}{M}\left(\frac{1}{2}+\ln\frac{\epsilon q}{2}\right),\label{eq:D8}
\end{equation}
we find
\begin{multline}
\frac{\hbar^{2}q^{2}}{M}\int_{r<\epsilon}d{\bf r}\left[r\ln\frac{qr}{2}\Omega_{1}^{(\sigma^{\prime})*}\left(\varphi\right)\right]\left[\frac{\delta\left(r\right)}{r^{2}}\Omega_{1}^{(\sigma)}\left(\varphi\right)\right]\\
=\delta_{\sigma\sigma^{\prime}}\lim_{\Lambda\rightarrow\infty}\left[-\frac{\hbar^{2}q^{2}}{2M}\left(2\gamma-1\right)-\frac{\hbar^{2}q^{2}}{M}\ln\frac{\Lambda}{q}-8\pi^{2}q^{2}\int_{\Lambda}^{\infty}\frac{kdk}{\left(2\pi\right)^{2}}\frac{\hbar^{2}k^{2}}{2M}\frac{1}{k^{4}}\right],\label{eq:D9}
\end{multline}
and then
\begin{multline}
\frac{\pi\hbar^{2}q}{M}\int_{r<\epsilon}d{\bf r}\left[N_{1}\left(qr\right)\Omega_{1}^{(\sigma^{\prime})*}\left(\varphi\right)\right]\left[\frac{\delta\left(r\right)}{r^{2}}\Omega_{1}^{(\sigma)}\left(\varphi\right)\right]\\
=\delta_{\sigma\sigma^{\prime}}\lim_{\Lambda\rightarrow\infty}\left[-\frac{\hbar^{2}\Lambda^{2}}{2M}-\frac{\hbar^{2}q^{2}}{M}\ln\frac{\Lambda}{q}-\int_{\Lambda}^{\infty}\frac{kdk}{\left(2\pi\right)^{2}}\frac{\hbar^{2}k^{2}}{2M}\left(\frac{8\pi^{2}}{k^{2}}+\frac{8\pi^{2}q^{2}}{k^{4}}\right)\right].\label{eq:D10}
\end{multline}
 As to the second term of Eq.(\ref{eq:D1}), we have
\begin{equation}
-\frac{\pi^{2}\hbar^{2}q^{4}}{4M}\int_{r<\epsilon}d{\bf r}\left[N_{1}\left(qr\right)\Omega_{1}^{(\sigma^{\prime})*}\left(\varphi\right)\right]\left[N_{1}\left(qr\right)\Omega_{1}^{(\sigma)}\left(\varphi\right)\right]=-\frac{\hbar^{2}q^{2}}{M}\left[\int_{r<\epsilon}d{\bf r}\frac{\Omega_{1}^{(\sigma^{\prime})*}\left(\varphi\right)}{r}\frac{\Omega_{1}^{(\sigma)}\left(\varphi\right)}{r}+O\left(\epsilon q\right)^{2}\right].\label{eq:D11}
\end{equation}
 Using the Fourier transform of $\Omega_{1}^{(\sigma)}\left(\varphi\right)/r$,
i.e., 
\begin{equation}
\mathcal{F}\left[\frac{\Omega_{1}^{(\sigma)}\left(\varphi\right)}{r}\right]=-i\sqrt{2\pi}\frac{e^{i\sigma\varphi_{{\bf k}}}}{k},\label{eq:D12}
\end{equation}
 we obtain
\begin{align}
 & -\frac{\pi^{2}\hbar^{2}q^{4}}{4M}\int_{r<\epsilon}d{\bf r}\left[N_{1}\left(qr\right)\Omega_{1}^{(\sigma^{\prime})*}\left(\varphi\right)\right]\left[N_{1}\left(qr\right)\Omega_{1}^{(\sigma)}\left(\varphi\right)\right]\nonumber \\
= & -\delta_{\sigma\sigma^{\prime}}\frac{4\pi^{2}\hbar^{2}q^{2}}{M}\int\frac{kdk}{\left(2\pi\right)^{2}}\frac{1-J_{0}\left(k\epsilon\right)}{k^{2}}\nonumber \\
= & -\delta_{\sigma\sigma^{\prime}}\frac{4\pi^{2}\hbar^{2}q^{2}}{M}\lim_{\Lambda\rightarrow\infty}\left[\int_{0}^{\Lambda}\frac{kdk}{\left(2\pi\right)^{2}}\frac{1-J_{0}\left(k\epsilon\right)}{k^{2}}+\int_{\Lambda}^{\infty}\frac{kdk}{\left(2\pi\right)^{2}}\frac{1-J_{0}\left(k\epsilon\right)}{k^{2}}\right]\nonumber \\
= & \delta_{\sigma\sigma^{\prime}}\lim_{\Lambda\rightarrow\infty}\left[-\frac{\hbar^{2}q^{2}}{M}\left(\gamma+\ln\frac{\epsilon\Lambda}{2}\right)-\int_{\Lambda}^{\infty}\frac{kdk}{\left(2\pi\right)^{2}}\frac{\hbar^{2}k^{2}}{2M}\frac{8\pi^{2}q^{2}}{k^{4}}\right]\label{eq:D13}
\end{align}
in the limit $\epsilon q\rightarrow0$. Combining Eqs.(\ref{eq:D10})
and (\ref{eq:D13}), we finally obtain Eq.(\ref{eq:3.19}) in the
maintext, i.e.,
\begin{multline}
\frac{\pi^{2}\hbar^{2}q^{2}}{4M}\int_{r<\epsilon}d{\bf r}N_{1}\left(qr\right)\Omega_{1}^{(\sigma^{\prime})*}\left(\varphi\right)\nabla_{{\bf r}}^{2}\left[N_{1}\left(qr\right)\Omega_{1}^{(\sigma)}\left(\varphi\right)\right]\\
=\delta_{\sigma\sigma^{\prime}}\lim_{\Lambda\rightarrow\infty}\left[-\frac{\hbar^{2}\Lambda^{2}}{2M}-\frac{\hbar^{2}q^{2}}{M}\ln\frac{\Lambda}{q}-\frac{\hbar^{2}q^{2}}{M}\left(\gamma+\ln\frac{\epsilon\Lambda}{2}\right)-\int_{\Lambda}^{\infty}\frac{kdk}{\left(2\pi\right)^{2}}\frac{\hbar^{2}k^{2}}{2M}\left(\frac{8\pi^{2}}{k^{2}}+\frac{16\pi^{2}q^{2}}{k^{4}}\right)\right].\label{eq:D14}
\end{multline}

In the follows, let us look at Eq.(\ref{eq:3.21}). It is easily found
\begin{equation}
{\bf I}_{2}^{(\sigma\sigma^{\prime})}=\int_{r<\epsilon}d{\bf r}\frac{\Omega_{1}^{(\sigma^{\prime})*}\left(\varphi\right)}{r}\frac{\Omega_{1}^{(\sigma)}\left(\varphi\right)}{r}+O\left(\epsilon q\right)^{2}.\label{eq:D15}
\end{equation}
Using the Fourier transform of $\Omega_{1}^{(\sigma)}\left(\varphi\right)/r$,
i.e., Eq.(\ref{eq:D12}), we have
\begin{equation}
{\bf I}_{2}^{(\sigma\sigma^{\prime})}=\delta_{\sigma\sigma^{\prime}}4\pi^{2}\int\frac{kdk}{\left(2\pi\right)^{2}}\frac{1-J_{0}\left(k\epsilon\right)}{k^{2}}\label{eq:D16}
\end{equation}
in the limit $\epsilon q\rightarrow0$, and then ${\bf I}_{2}^{(\sigma\sigma^{\prime})}$
can alternatively be written as
\begin{eqnarray}
{\bf I}_{2}^{(\sigma\sigma^{\prime})} & = & \delta_{\sigma\sigma^{\prime}}\lim_{\Lambda\rightarrow\infty}\left[4\pi^{2}\int_{0}^{\Lambda}\frac{kdk}{\left(2\pi\right)^{2}}\frac{1-J_{0}\left(k\epsilon\right)}{k^{2}}+4\pi^{2}\int_{\Lambda}^{\infty}\frac{kdk}{\left(2\pi\right)^{2}}\frac{1-J_{0}\left(k\epsilon\right)}{k^{2}}\right]\nonumber \\
 & = & \delta_{\sigma\sigma^{\prime}}\lim_{\Lambda\rightarrow\infty}\left[\gamma+\ln\frac{\epsilon\Lambda}{2}+4\pi^{2}\int_{\Lambda}^{\infty}\frac{kdk}{\left(2\pi\right)^{2}}\frac{1}{k^{2}}\right],\label{eq:D17}
\end{eqnarray}
 where we have used
\begin{equation}
\lim_{\Lambda\rightarrow\infty}\int_{\Lambda}^{\infty}\frac{kdk}{\left(2\pi\right)^{2}}\frac{J_{0}\left(k\epsilon\right)}{k^{2}}=0.\label{eq:D18}
\end{equation}

\end{widetext}

\end{document}